\documentclass[twocolumn, usenatbib]{mnras}
\usepackage{savesym}
\usepackage{graphicx}
\usepackage{longtable}
\usepackage{changepage}

\expandafter\let\csname equation*\endcsname\relax
  \expandafter\let\csname endequation*\endcsname\relax 
\usepackage{subfig}
\usepackage{amsmath}
\usepackage{amssymb}
\usepackage{verbatim}
\usepackage[yyyymmdd,hhmmss]{datetime}
\usepackage{array}
\usepackage{times}
\usepackage{xcolor}

\DeclareRobustCommand{\DE}[3]{#2}
\let\DEthebibliography\thebibliography
\def\thebibliography{\DeclareRobustCommand{\DE}[3]{##3}\DEthebibliography}

\newcommand\ntde{{62}}

\title[ TDE luminosity functions from first principles  ]{ The optical, UV-plateau and X-ray tidal disruption event luminosity functions reproduced from first principles  }
\author [Andrew Mummery, Sjoert van Velzen]{Andrew Mummery$^1$\thanks{E-mail:
andrew.mummery@physics.ox.ac.uk},  Sjoert van Velzen$^2$
\\
$^1$Oxford Theoretical Physics, Beecroft Building,  Clarendon Laboratory, Parks Road, Oxford, OX1 3PU, United Kingdom\\
$^2$Leiden Observatory, Leiden University, Postbus 9513, 2300 RA Leiden, The Netherlands
}

\date{}

\begin{document}

\pagerange{\pageref{firstpage}--\pageref{lastpage}} \pubyear{2024}

\maketitle

\label{firstpage}

\begin{abstract} 
We reproduce the luminosity functions of the early-time peak optical luminosity, the late-time UV plateau luminosity, and the peak X-ray luminosity of tidal disruption events, using an entirely first-principles theoretical approach. We do this by first fitting three free parameters of the tidal disruption event black hole mass distribution using the observed distribution of late time  UV plateau luminosities, using a time-dependent relativistic accretion model. Using this black hole mass distribution we are then, with no further free parameters of the theory, able to reproduce exactly the peak X-ray luminosity distribution of the tidal disruption event population. This proves that the X-ray luminosity of tidal disruption events are sourced from the same accretion flows which produce the late time UV plateau.  Using an empirical scaling relationship between peak optical luminosities and black hole masses, itself calibrated using the same relativistic accretion theory, we are able to reproduce the observed peak optical luminosity function, again with no additional free parameters. Implications of these results include that there is no tidal disruption event ``missing energy problem'',  that the optical and X-ray selected tidal disruption event populations are drawn from the same black hole mass distribution, that the early time optical luminosity in tidal disruption events is somewhat simple, at least on the population level, and that future LSST observations will be able to constrain the black hole mass function at low masses. 
\end{abstract}

\noindent

\section{Introduction}
Tidal disruption events occur in the centres of (previously) quiescent  galaxies when an unfortunate star is scattered by $N$-body gravitational interactions onto a near-radial orbit about the central supermassive black hole. Once the tidal force of the black hole's spacetime overcomes the star's self gravity, the star is unbounded and becomes a debris stream which ultimately rains down onto the black hole powering bright flares observed across the electromagnetic spectrum. 

A ``canonical'' tidal disruption event is typically discovered as a bright flare in optical bands, which may or may not be associated with a rapid rise in X-ray flux, from the centre of a galaxy. Typical optical luminosities peak at around $L \sim 10^{43-44}$ erg/s, as do the brightest X-ray flares, before decaying away on $\sim$ month timescales. On significantly longer timescale ($\sim$ years) a pronounced plateau is observed in the optical/UV  light curves \citep{vanVelzen19, MumBalb20a} which is most easily observed in the UV \citep{vanVelzen19} but can also be detected in the optical \citep{Mummery_et_al_2024}. This plateau is associated with a transition from the (currently poorly understood) physics of the early-time optical flare to a disc-dominated state \citep{MumBalb20a, Mummery_et_al_2024}.

While a rare event in any one galaxy \citep[typical rates are of order $\sim 10^{-4}$ galaxy$^{-1}\,$ year$^{-1}$][]{Rees88, Magorrian99}, all sky surveys in the optical \citep[e.g., the ASASSN][and ZTF \citealt{vanVelzen19b} surveys]{Holoien16b} and X-ray \citep[such as the XMM slew][and eRosita \citealt{Sazonov21} surveys]{Esquej08},  have lead to a growing number of discovered systems, a population which now stands at roughly 100 individual sources. These hard-won data sets mean that it now possible to pose questions regarding the  properties of the populations of these exotic objects.  This is an important step-change in tidal disruption event science, as while the properties of some individual sources are well described by black hole accretion models \citep[e.g. the well-behaved source ASASSN-14li][]{MumBalb20a, Wen20}, other individual sources show more extreme properties \citep[such as large amplitude short-timescale variability e.g.,][]{Wevers21}.  It is an important test of theoretical models of tidal disruption event evolution that the gross population-level properties of these systems can be reproduced with physically reasonable parameters.

Recently \cite{Mummery_et_al_2024} demonstrated that the late-time UV plateau luminosity observed from tidal disruption events offer a particularly promising route to understanding the properties of the population of tidal disruption events at large. The reason for this is that the amplitude of this late-time luminosity is strongly correlated with the black hole mass at the centre of the event $L_{\rm plat} \propto M_\bullet^{2/3}$, meaning that the distribution of plateau luminosities observed in these systems might offer a route into constraining the black hole mass distribution of these events, a key step in constraining the wider (e.g., peak optical and X-ray luminosity) observed distributions of these systems. \cite{Mummery_et_al_2024} demonstrated that the masses inferred from the late time plateau  are consistent with the known $M_\bullet - \sigma$ relationship extrapolated down from higher mass galaxies, which is a good start.

\begin{figure}
    \centering
    \includegraphics[width=\linewidth]{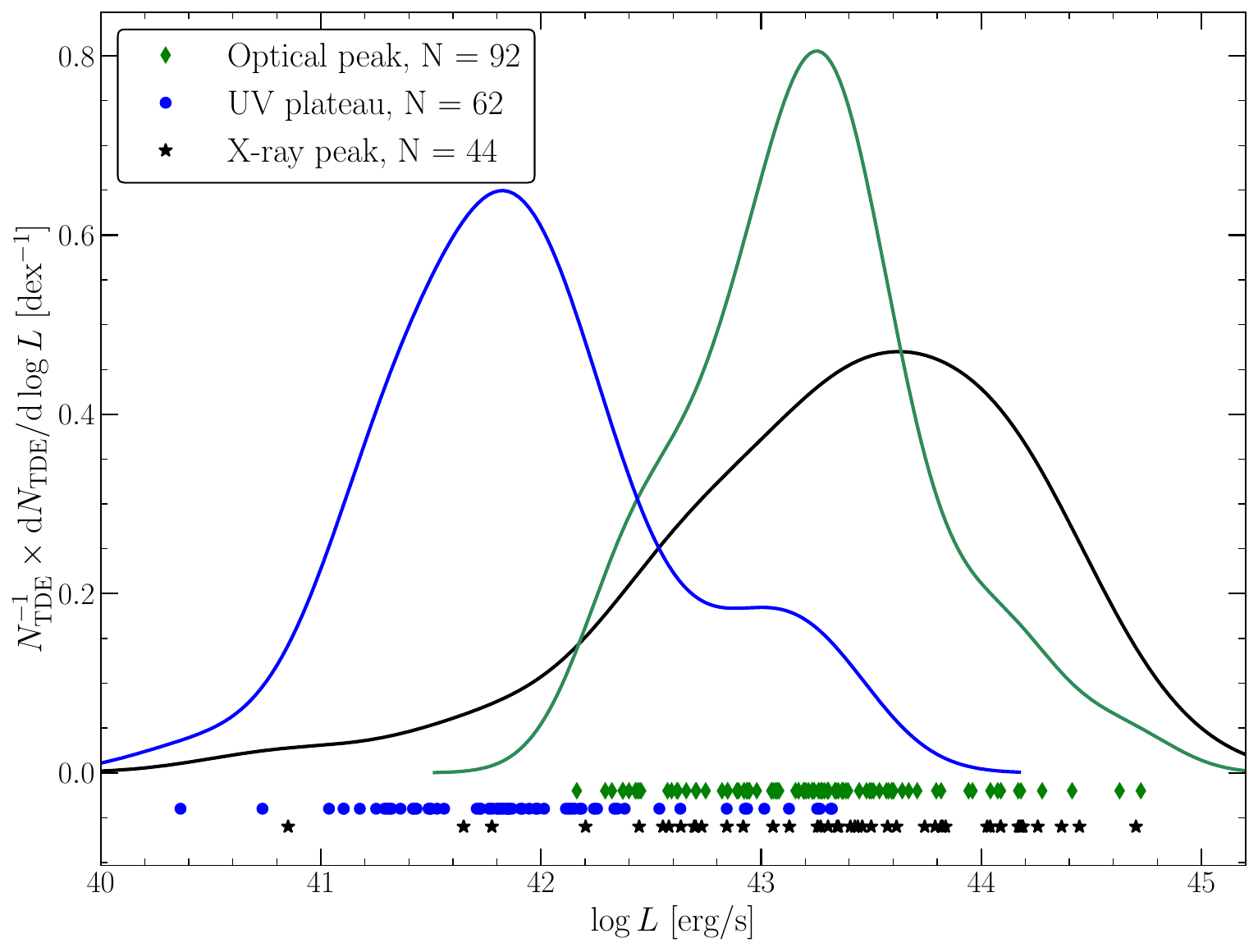}
    \includegraphics[width=\linewidth]{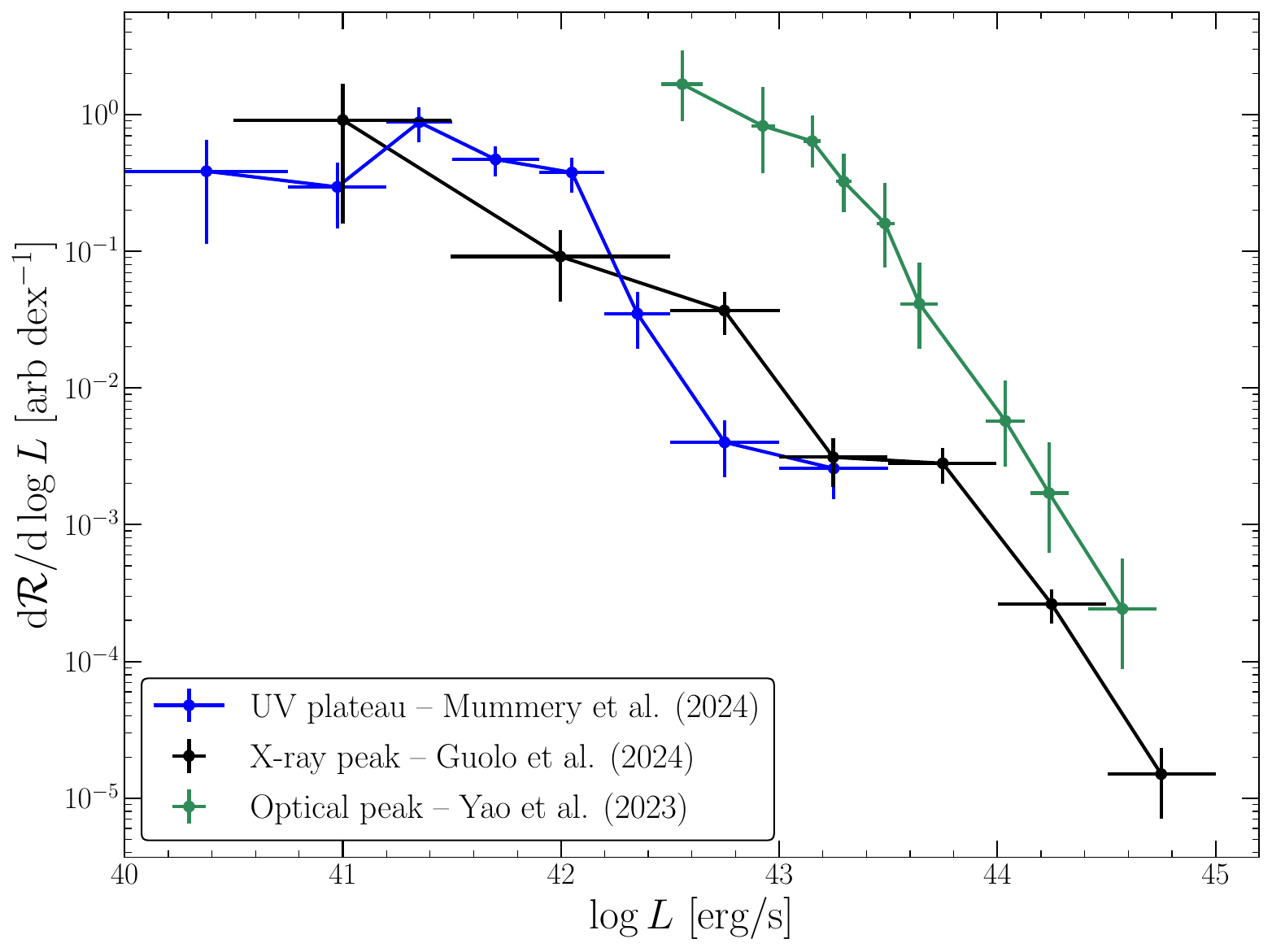}
    \caption{The observational data which we seek to reproduce in this work.  Upper: the observed tidal disruption event luminosity distributions in optical (early time), ultraviolet (late time) and X-ray (peak) bands. The smooth curves are computed using Gaussian Kernel Density Estimation, while individual data points are shown below each curve. The structure of each luminosity distribution shows unique features. Lower: the three corresponding luminosity functions (normalised so that each curve integrates to unity). The different structures of each luminosity distribution are most noticeable when expressed in terms of intrinsic rates.      }
    \label{fig:thedists}
\end{figure}

Concurrently, a number of observational surveys have lead to tight constraints being placed on the intrinsic rate at which the early-time optical \cite{Yao23}, late-time UV \cite{Mummery_et_al_2024} and X-ray \cite{Guolo24} luminosities of tidal disruption events are distributed\footnote{In this paper we use the symbol $L_{g}$ ($L_{UV}$) to represent $\nu L_\nu$ for narrow-band observations in the $g$-band (UV bands), and $L_X$ for the integrated luminosity across the broader $0.3-10$ keV X-ray bandpass.}. Current observational data constrains both the observed distributions (Figure \ref{fig:thedists}; data from \cite{Yao23}, \cite{Mummery_et_al_2024} and \cite{Guolo24}) but also the relative rates of certain properties in the wider universe (i.e., the above distributions once Malmquist biases are corrected for;  lower panel of Figure \ref{fig:thedists}).  

The simplest theoretical question that can be asked is whether any one model can reproduce, from first principles, the observations summarised in Figure \ref{fig:thedists}. This question has clear physical content: the distributions and rates have fundamentally different structures (particularly when plotted as intrinsic rates) from observational band to observational band. This rules out naive models which (for example) scale each luminosity with the Eddington limit of the host black hole, as this would imply a simple shifting of the same distribution.    As far as the authors are aware this question has not yet been posed in the literature. It is the purpose of this paper to demonstrate that all three distributions can be reproduced entirely with results derived from time-dependent relativistic accretion theory.

The layout of this paper is as follows. In section \ref{sec:moretdes} we introduce the TDE sample used in this work, which includes an extended UV plateau population.   In section \ref{sec:framework} we introduce the underlying theoretical framework used in this paper, and review the accretion models introduced in \cite{Mummery_et_al_2024}. In section \ref{sec:results} we present the results of our analysis, before discussing their implications in section \ref{sec:implications}. We conclude in section \ref{sec:conclusions}. A detailed description of the statistical formalism used in constraining the parameters of the black hole mass distribution is presented in Appendix \ref{app:stats}. 

\section{The observed TDE population}\label{sec:moretdes}
In this work we use the {\tt manyTDE}\footnote{All of this data is publicly available at \url{https://github.com/sjoertvv/manyTDE.}} database of tidal disruption event lightcurve parameters. This data set was compiled during the work presented in \cite{Mummery_et_al_2024}, although since this publication more TDEs have been discovered, and some of these sources now have detectable late-time plateaus. We summarise the properties of the 14 new TDEs with detected late-time plateaus in Table \ref{mass_plat_table} (see the tables in Appendix \ref{app:discovery} for discovery references for each new TDE, and galactic properties). The remainder of our optical/UV sample is the dataset presented in \cite{Mummery_et_al_2024}, although more data has been collected since publication and some of these older TDEs therefore now have subtly different plateau luminosity measurements.  

The data reduction, and the process of plateau luminosity measuring, for each of these new TDEs follows exactly the procedure spelled out in \cite{Mummery_et_al_2024}, their section 5. With the UV plateau luminosity extracted, we can use the plateau luminosity-black hole mass scaling relationship to provide estimates of the black hole masses of each of these new TDEs. These masses are also presented (with $1\sigma$ uncertainties) in Table \ref{mass_plat_table}. 

\begin{figure}
    \centering
    \includegraphics[width=\linewidth]{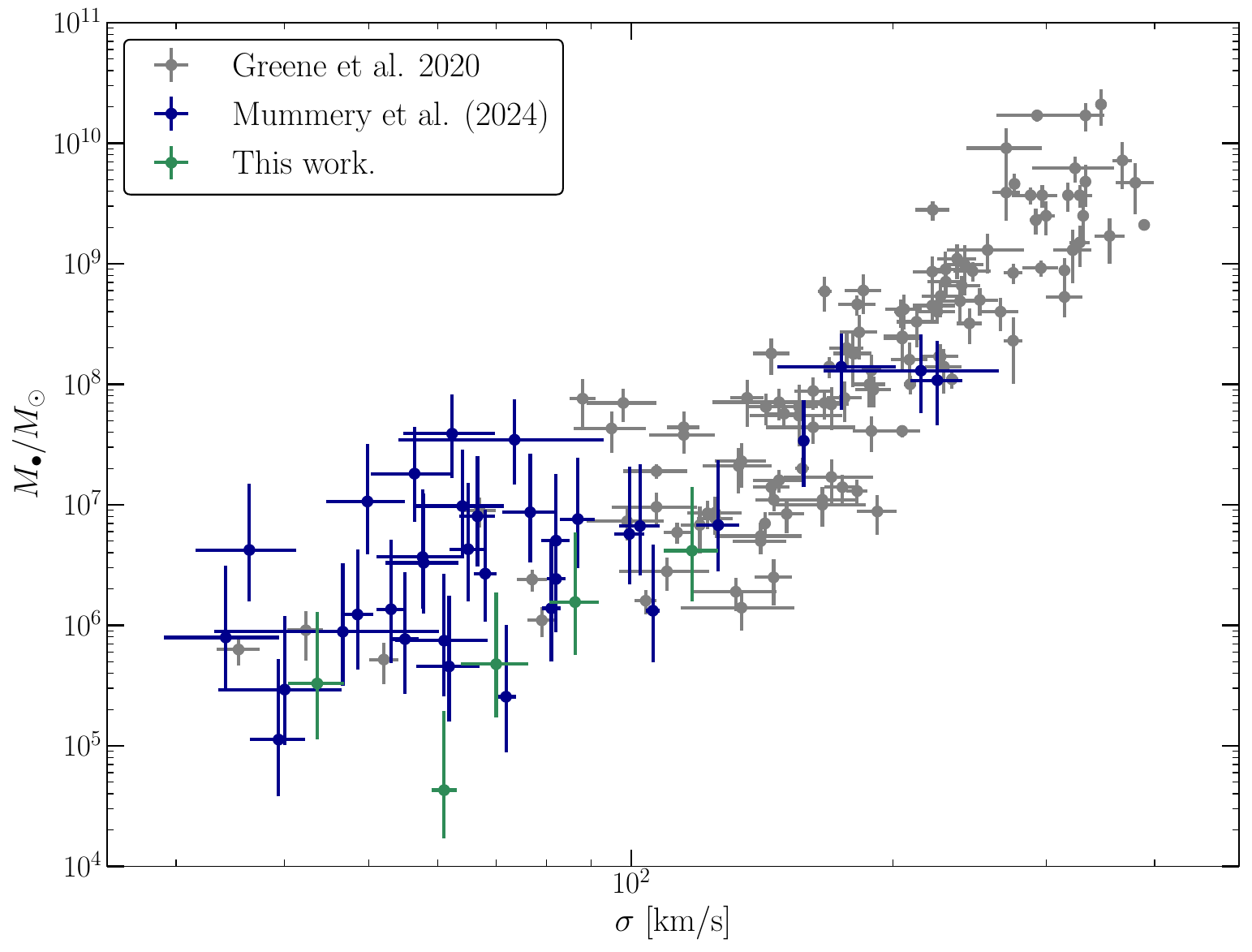}
    \includegraphics[width=\linewidth]{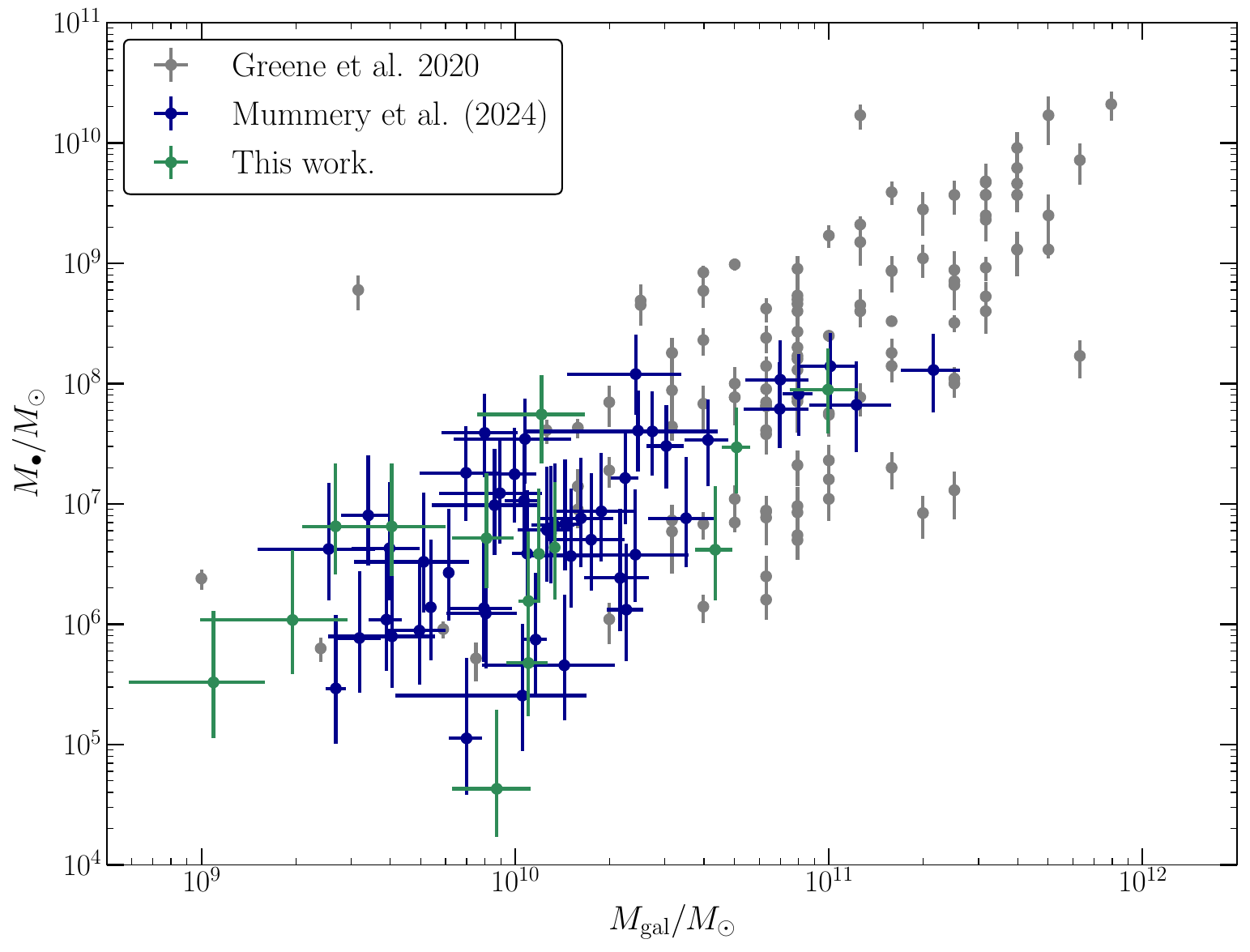}
    \caption{Updated $M_\bullet-\sigma$ (upper panel) and $M_\bullet-M_{\rm gal}$ (lower panel) scaling relationships including the new TDE measurements of Table \ref{mass_plat_table}. The (typically higher black hole mass) measurements are taken from the \citealt{Greene20} review, while the majority of the TDE population were first published in \citealt{Mummery_et_al_2024}. The new points remain consistent with the expected extrapolation of known relationships calibrated at higher masses.  }
    \label{fig:new_scaling}
\end{figure}

Many of the TDEs with plateau-inferred black hole masses also have independent estimates of their black hole masses from galactic scaling relationships, namely the velocity dispersion $\sigma$ and galaxy mass $M_{\rm gal}$ relationships. In Figure \ref{fig:new_scaling} we place the 14 new TDEs (not every TDE has a velocity dispersion measurement) onto the $M_\bullet-\sigma$ and $M_\bullet-M_{\rm gal}$ scaling relationships of \cite{Greene20}. The new points remain consistent with the expected extrapolation of known relationships calibrated at higher masses. Note the systematically lower masses of the $M_\bullet-\sigma$ measurements of the newer population, this may be an example of Malmquist bias -- those TDEs with fainter plateaus (a result of lower black hole masses) need more data to detect, and will in general be detected later in a survey. 

In addition we use the peak X-ray luminosity database collected in \cite{Guolo24}, and the peak $g$-band luminosity function presented in \cite{Yao23}. We have verified that the  luminosity function presented in \cite{Yao23} is consistent with that which would be inferred from our extended {\tt manyTDE} dataset. 

\renewcommand{\arraystretch}{1.25}
\begin{table}
\centering
\begin{tabular}{ |p{2.0cm}|p{2.0cm}|p{2.0cm}|  }
\hline
TDE Name & $\log_{10} L_{\rm plat}$ & $\log_{10} M_\bullet$  \\
& $ \log_{10} {\rm erg\, s^{-1}}$ & $ \log_{10} M_\odot$  \\
\hline
 AT2017eqx & $41.98^{+0.20}_{-0.12}$ & $6.81^{+ 0.52}_{- 0.40}$ \\ \hline 
 AT2018bsi & $41.83^{+0.13}_{-0.13}$ & $6.62^{+ 0.53}_{- 0.42}$ \\ \hline 
 AT2019eve & $41.42^{+0.04}_{-0.06}$ & $6.04^{+ 0.58}_{- 0.45}$ \\ \hline 
 AT2020ksf & $41.91^{+0.05}_{-0.05}$ & $6.71^{+ 0.54}_{- 0.42}$ \\ \hline 
 AT2020vdq & $41.04^{+0.07}_{-0.07}$ & $5.52^{+ 0.59}_{- 0.46}$ \\ \hline 
 AT2021ack & $41.85^{+0.11}_{-0.09}$ & $6.64^{+ 0.54}_{- 0.43}$ \\ \hline 
 AT2021gje & $43.13^{+0.04}_{-0.03}$ & $7.95^{+ 0.34}_{- 0.36}$ \\ \hline 
 AT2021lo & $42.84^{+0.05}_{-0.04}$ & $7.74^{+ 0.33}_{- 0.41}$ \\ \hline 
 AT2022dbl & $41.18^{+0.04}_{-0.03}$ & $5.68^{+ 0.60}_{- 0.44}$ \\ \hline 
 AT2022bdw & $41.53^{+0.06}_{-0.05}$ & $6.19^{+ 0.58}_{- 0.44}$ \\ \hline 
 AT2022rz & $41.98^{+0.11}_{-0.10}$ & $6.81^{+ 0.52}_{- 0.41}$ \\ \hline 
 AT2022hvp & $42.54^{+0.04}_{-0.04}$ & $7.47^{+ 0.33}_{- 0.38}$ \\ \hline 
 AT2022dsb & $41.82^{+0.02}_{-0.02}$ & $6.58^{+ 0.54}_{- 0.40}$ \\ \hline 
 AT2023clx & $40.36^{+0.12}_{-0.11}$ & $4.63^{+ 0.65}_{- 0.40}$ \\ \hline 
\end{tabular}
\caption{The black hole masses of the 14 new TDEs in our sample, computed from the (\citealt{Mummery_et_al_2024}) plateau luminosity scaling relationship. The quoted error ranges correspond to $1\sigma$ uncertainties. The plateau luminosity here is measured at $\nu = 10^{15}$ Hz. }
\label{mass_plat_table}
\end{table}

\section{Analysis framework }\label{sec:framework}
It is the aim of this paper to reproduce the population-level properties of tidal disruption events as observed at optical, ultraviolet and X-ray frequencies.  Theoretically, the reproduction of a distribution of ``observables'' is a fundamentally rather simple two step process. First,  one needs to specify the intrinsic distribution of physical tidal disruption event ``input'' parameters (in this case the distributions of  e.g., the black hole masses and spins, the stellar masses, etc., which are involved in these events). The second step  is then turning each set of input parameters into a predicted ``output'' observable (in our case these outputs are observed luminosities) using a physical model. In this section we begin by introducing the physical model, then we discuss the functional form of the ``input'' parameter distributions considered in this work, before concluding with a discussion of the raw data we are aiming to reproduce.

\subsection{The relativistic accretion flow model }
The fundamental physical model used in this work is the time-dependent relativistic accretion flow model which was developed in \cite{MumBalb20a}, and extended in \cite{Mummery_et_al_2024}. This model is discussed in significant detail in the code paper \cite{mummery2024fitted}, and in this paper we therefore only give a broad summary of its constituent components, and refer the reader to these earlier papers if more detail is required.  

The model makes the assumption that one half of the disrupted stellar debris ultimately circularises into an accretion flow, with initial radial scale given by the circularisation radius of the disruption\footnote{The circularisation radius is the radius at which one half of the stellar debris would settle if its initial parabolic orbit was transformed into a circular orbit while conserving angular momentum.}. This disc material thereafter evolves according to the time-dependent general relativistic thin disc equation \citep{EardleyLightman75, Balbus17}, which is the relativistic analogue of the classical Lynden-Bell and Pringle Newtonian disc equation \citep{LBP74}.  This governing equation evolves the disc density as a function of radius and time, from which the locally liberated photon flux of the disc can be computed. This local photon flux is assumed to be emitted quasi-thermally, with a temperature dependent colour-correction factor \citep{Done12} included to model disc opacity effects. For the sake of computational speed we use the analytical disc solutions of \cite{Mummery23a}, which assume a vanishing ISCO stress for simplicity.  This allows a rest-frame spectral energy distribution of the photons emitted from the accretion flow to be computed at every time post-disruption. The observer is assumed to be located at a distance $D$ from the disc, orientated at an angle $\theta_{\rm obs}$ from the black hole's spin axis. The observed spectral energy distribution of the emitted radiation is then computed using full relativistic ray-tracing calculations. We use a modified \citep{Ingram19, Mummery_et_al_2024} version of the publicly available code {\tt YNOGK} \cite{YangWang13}, which is based on {\tt GEOKERR} \cite{DexterAgol09} for the ray tracing calculations. This observed spectral energy distribution can then be converted into observer frame light curves, which incorporate all relativistic effects. For a given set of input parameters, namely the black hole mass and spin $M_\bullet, a_\bullet$; the initial disc mass $M_{\rm disc}$ (taken to be one half of the incoming stellar mass); the  ``viscous'' evolutionary timescale of the disc $t_{\rm visc}$, and observer inclination $\theta_{\rm obs}$, disc light curves can be produced at all frequency bands. 

\subsection{Physical parameter distributions} 
As discussed above, once a set of input parameters are specified, the output (luminosity at a given frequency versus time) of the disc model can be computed. At first glance it may appear however that the parameter space available to fitting observed luminosity functions will be incredibly broad, and the problem will likely be under-constrained. Fortunately, however, nearly all of the key parameter distributions are known (to good approximation) a priori, as discussed in \cite{Mummery_et_al_2024}. In this work we fix the parameter distributions of all parameters except the black hole mass to that used in \cite{Mummery_et_al_2024}. As an explicit  example, the stellar mass distribution is fixed to that of the \cite{Kroupa01} initial stellar mass function, multiplied by the \cite{Wang04} intrinsic stellar mass dependence of the tidal disruption event rate. We use the \cite{Kippenhahn90} stellar mass-radius relationship throughout this work, so that only the stellar mass is a free parameter for a given disruption. We do not include any contribution to the tidal disruption event rate from red giants or white dwarfs.  Other parameters are similarly simply distributed, we take flat (uninformative) prior distributions on the black hole spin parameter $a_\bullet$ for each black hole. We further assume that the observing inclination $\cos i$, and incoming stellar orbit angle $\sin \phi_{\rm orb}$ are uniformly distributed, as would be the case for an isotropic observer distribution (in the case of $\theta_{\rm obs}$) or isotropic galactic centre stellar population (in the case of $\phi_{\rm orb}$).  In our sample we only include full disruptions (i.e., those with penetration factors $\beta \geq 1$), and sample the impact parameters according to the prescription of \cite{Stone16} \citep[see][for more details]{Mummery_et_al_2024}. 

To leading order therefore we have only one parameter distribution which must be fit from the data, the distribution of black hole masses in observed tidal disruption events. In this work we parameterise the tidal disruption event black hole mass distribution with the following functional form \citep{Schechter76}
\begin{equation}
    p_{M_\bullet}(m_\bullet) \propto  {m_\bullet^{\alpha_h} \over  1 + (M_c/m_\bullet)^{\alpha_l-\alpha_h} } \exp\left[- \left({m_\bullet\over M_g}\right)^\gamma\,\right] .
\end{equation}
This  represents a broken power-law with a high mass exponential cut off. At low black hole masses $(M_\bullet \ll M_c)$ this function is well approximated by a single power law with index $p_{M_\bullet} \sim m_\bullet^{\alpha_l}$, while for higher masses $(M_c \ll M_\bullet \ll M_g)$ the power-law slope becomes $p_{M_\bullet} \sim m_\bullet^{\alpha_h}$, with the transition scale occurring at black hole masses $M_\bullet \sim M_c$. The very high mass behaviour is that of an exponential decay owing physically to the scarcity of high mass galaxies. We take $M_g = 6.4\times 10^7 M_\odot$ and $\gamma = +0.49$, which are the values presented in \cite{Shankar04}. Our results are only weakly sensitive to the values of $\gamma$ and $M_g$, as the high mass tidal disruption event rate is principally controlled by Hills mass suppression (discussed in detail in Appendix \ref{app:stats}), which leads to a super-exponential rate suppression, rather than this exponential suppression. The aim of this analysis is to infer the parameters $\alpha_l, \alpha_h$ and $M_c$ from tidal disruption event observations.

\subsection{The optical/UV data}
As was demonstrated in \cite{Mummery_et_al_2024}, the properties of the late-time UV plateau observed in tidal disruption event light curves are well described by the relativistic accretion theory summarised above. It is this data set therefore which we use to constrain the parameters of the  black hole mass distribution introduced above. 

As discussed in section \ref{sec:moretdes},  for the raw data we use the compilation of tidal disruption event late time UV plateau luminosities presented in \cite{Mummery_et_al_2024}, with the addition of 14 new TDEs discovered since publication of this work.  For each tidal disruption event with a late time plateau we also have a measured maximum optical luminosity, which will be used in later sections to compute the maximum observable volume of each source.   The raw luminosity data is presented in Figure \ref{fig:thedists}, as an observed probability density function in the upper panel, and as an intrinsic rate in the lower panel. The intrinsic rate is computed by weighting the probability density function by the inverse of the observable volume of each source (see Appendix \ref{app:stats} for more details), so that harder to detect sources get a larger weight and are more accurately represented (i.e., this procedure accounts for Malmquist bias). The uncertainty in each rate value is given by the Poisson noise for the number of tidal disruption event sources which make up that bin.

In Appendix \ref{app:stats} we construct a likelihood (denoted ${\cal L}$) which describes the probability of observing a given distribution of tidal disruption event UV plateau luminosities, given a set of black hole mass distribution parameters $\left\{ \xi \right\} = (\alpha_l, \alpha_h, M_c)$. This likelihood has the following functional form 
\begin{equation}
    \log {\cal L}\left(\left\{ \xi \right\}\right) = \sum_{j =1}^{N_{\rm TDE}} \, \log p_L\left(L_j; \left\{ \xi \right\} \right) , 
\end{equation}
where $L_j$ is the observed UV plateau luminosity of the $j^{\rm th}$ tidal disruption event in our sample.
We do not weight the likelihood for each TDE, and therefore fit to the raw observed plateau luminosity distribution, rather than the intrinsic distribution. Importantly, this means that the parameters of the black hole mass distribution we constrain are for the {\it observed} TDE black hole mass distribution of our flux-limited sample, rather than the {\it intrinsic} black hole mass function.  

The  computation of the plateau luminosity probability density function $p_L(l)$, given a set of black hole mass distribution parameters, is the key step in this analysis, and is spelt out in detail in Appendix \ref{app:stats}, to which we refer the interested reader.  Our sample comprises of $N_{\rm TDE} = \ntde$ sources with detected plateau luminosities. 

\subsection{Statistical tests of the fit}
We wish to contrast the predicted model luminosity functions with those which have been observed. In general there are two statistical properties of the  observed luminosity functions which  we wish to recover. These are the observed cumulative luminosity distribution function and the luminosity probability density function (which formally contain the same information), and of course  the intrinsic rate of the  observed luminosities (the luminosity function itself). So as to be precise moving forward, we spend some time here defining each of these distributions in turn.  The probability density function is defined by    
\begin{equation}\label{eq:pdf}
p(\log L) = {1 \over N_{\rm TDE}} {{\rm d} N_{\rm TDE} \over {\rm d} \log L}  ,
\end{equation}
where $L$ will represent either the UV plateau, optical peak or X-ray peak luminosity. In other words, the probability density function is the unweighted and normalised histogram of the observed luminosity data. The normalisation is simply defined so that 
\begin{equation}
\int_{-\infty}^{+\infty} p(\log L) \, {\rm d} \log L = 1. 
\end{equation}
This probability density function allows us to define the cumulative distribution function of the data $\Phi(\log L)$
\begin{equation}\label{eq:cdf}
\Phi(\log L) = \int_{-\infty}^{\log L} p\left(\log L'\right) \, {\rm d} \log L' .
\end{equation}
This is important because the cumulative distribution function allows non-parametric tests of goodness of fit to be performed, such as the two sample Kolmogorov-Smirnov (KS) test.

Finally the rate at which  a given luminosity in our sample appears intrinsically is given by 
\begin{equation}\label{eq:rate}
{{\rm d} {\cal R} \over {\rm d} \log L} \propto  {1 \over {\cal V}(\log L)} {{\rm d} N_{\rm TDE} \over {\rm d} \log L}  ,
\end{equation}
where ${\cal V}$ is the maximum observable volume associated with each tidal disruption event. 
 
The volume ${\cal V}$ can be simply estimated for  our sample as we know how each  tidal disruption event sample was constructed, namely from an optical survey telescope with a known flux detection limit, $F_{\rm lim}$ (although this analysis could also be performed for e.g., an X-ray instrument). In a Newtonian cosmology (which is a reasonable approximation for nearly all tidal disruption events which are found at $z < 0.1$), the maximum distance out to which a source with given peak luminosity (optical or X-ray)  $L_{{\rm peak}}$ can be observed is 
\begin{equation}
    D_{\rm max} \propto \sqrt{L_{{\rm peak}}} 
\end{equation}
and therefore an estimate of the volume out to which each tidal disruption event can be detected is (Appendix \ref{app:stats} for more discussion)
\begin{equation}
    {\cal V} \propto D^3_{\rm max} \propto \left(L_{{\rm peak}}\right)^{3/2} . 
\end{equation}

\begin{figure}
    \centering
    \includegraphics[width=\linewidth]{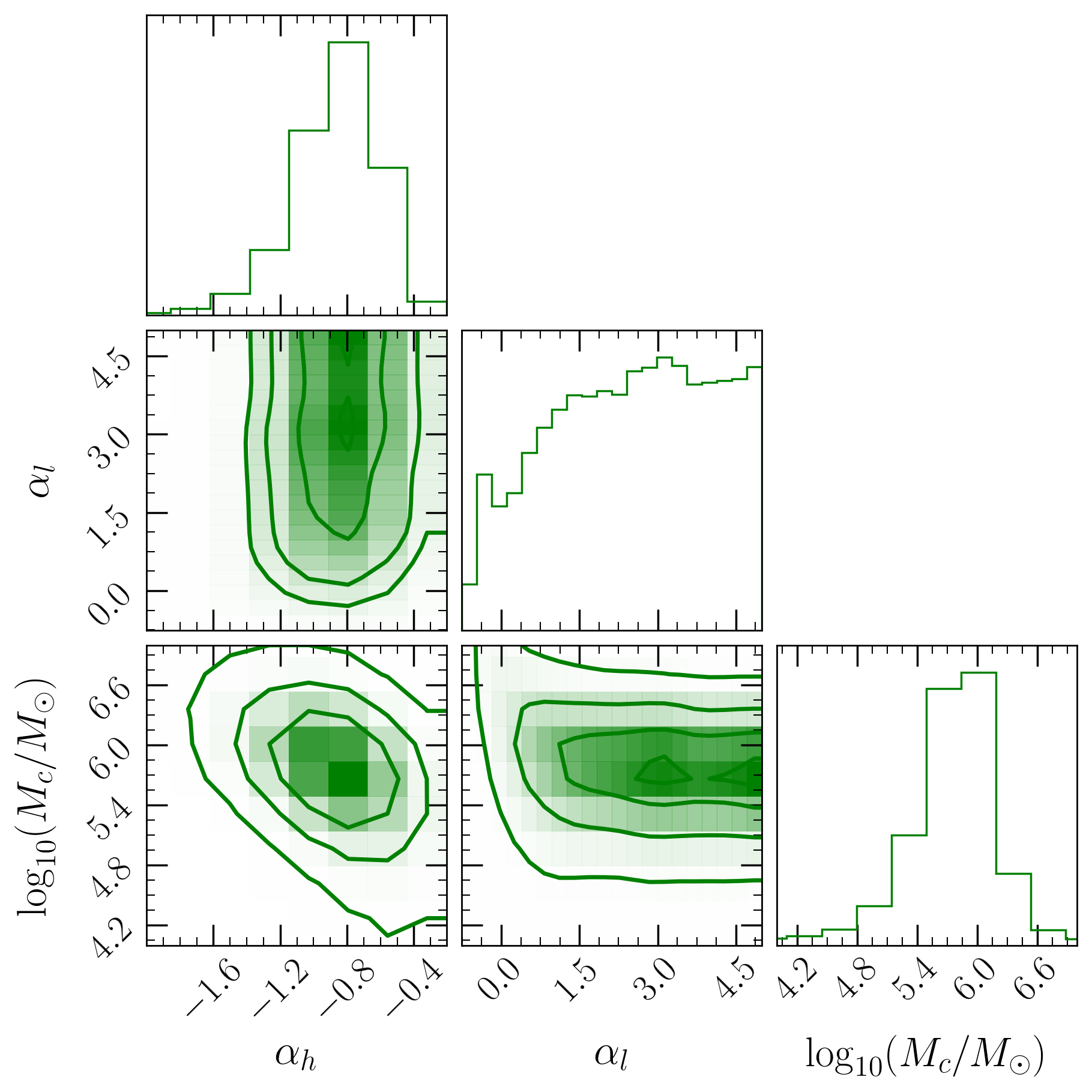}
    \caption{Posterior distributions of the black hole mass distribution parameters, resulting from a Monte Carlo Markov Chain analysis of the tidal disruption event UV plateau data. The data requires a turn-over in the tidal disruption event black hole mass rate at $M_\bullet \sim 10^6 M_\odot$ at high significance.  This  may be related to an intrinsic change in the black hole mass function, or could be driven by the lack of observing volume at these low luminosities, where current instruments do not typically have sensitivity. We cannot distinguish either scenario with current data. } 
    \label{fig:corner}
\end{figure}

\begin{figure}
    \centering
    \includegraphics[width=\linewidth]{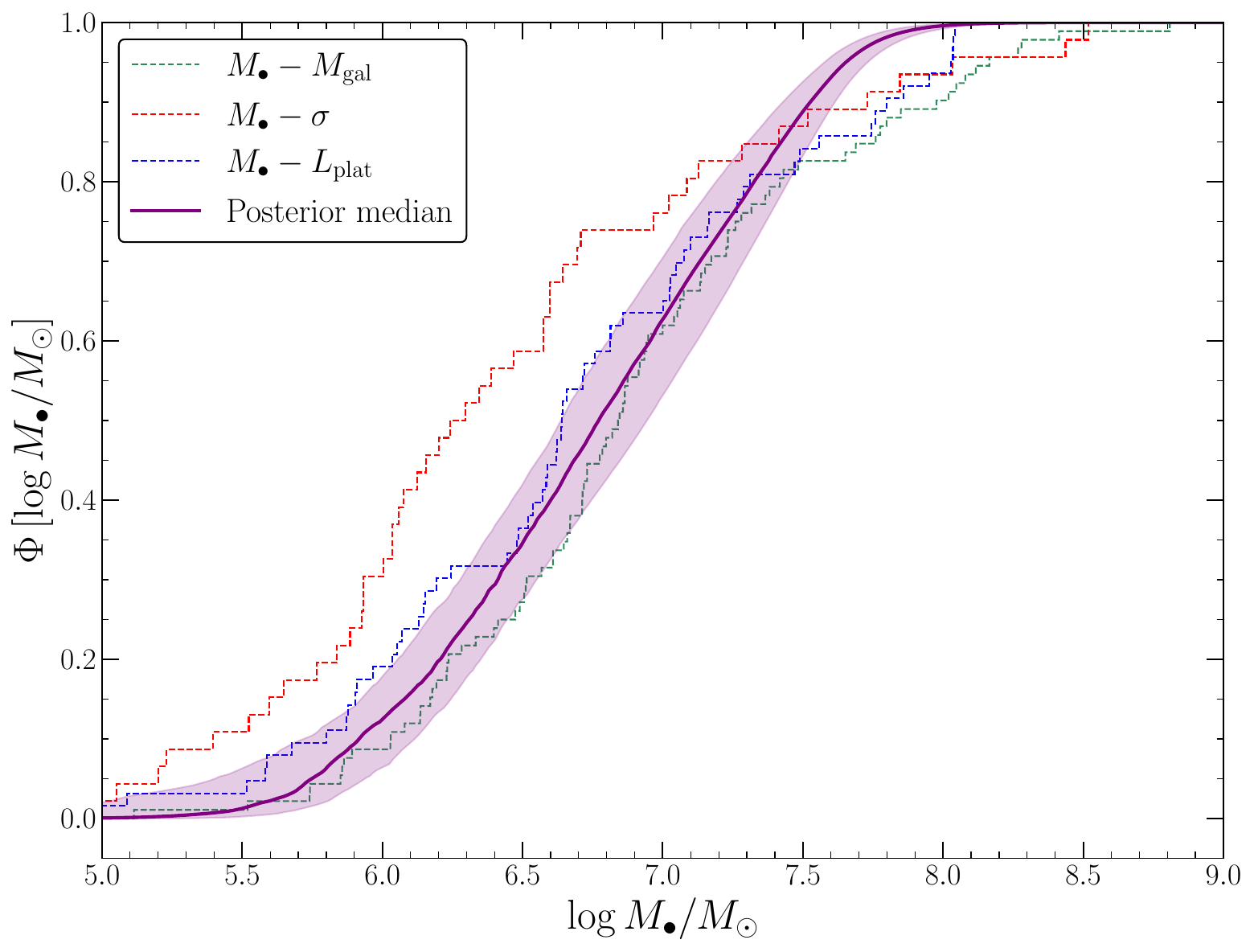}
    \caption{The cumulative distribution function of the observed TDE black hole masses of our sample, measured in three different ways. These are from the $M_\bullet-M_{\rm gal}$ relationship (green step function, \citealt{Greene20}), the $M_\bullet-\sigma$ relationship (red step function, \citealt{Greene20}), and the $M_\bullet-L_{\rm plat}$ scaling relationship (blue step function, \citealt{Mummery_et_al_2024}). The model  posterior median (purple curve) and $1 \sigma$ uncertainty (purple shaded region), are also shown, with the effects of Hills mass suppression included. The posterior median model produces a formally acceptable description of both the $M_\bullet-M_{\rm gal}$ and $M_\bullet-L_{\rm plat}$ populations, with the null hypothesis that the observed and simulated distributions are the same producing $p$-values $>0.1$ for a 2-sample KS test. The model is in contention with the $M_\bullet-\sigma$ population however ($p$-value = 0.001), see text for further discussion. }
    \label{fig:masscomp}
\end{figure}

\section{Results}\label{sec:results}
\subsection{Black hole mass distribution }
With the likelihood constructed, we can fit the three black hole mass distribution parameters $\alpha_l, \alpha_h$ and $M_c$ to the UV plateau data using standard techniques. To be precise, we run a Monte Carlo Markov Chain analysis, using the {\tt EMCEE} code \cite{EMCEE}. We use very broad, flat and uninformative priors on our fitting parameters, namely  $-5 < \alpha_{l}, \alpha_{h} < 5$, and $3 < \log M_c/M_\odot < 10$. The posterior distributions of these three parameters are shown in Figure \ref{fig:corner}. 

As is clear in Figure \ref{fig:corner}, this analysis is able to constrain both the high mass slope $\alpha_h$ and turnover mass $M_c$, of the tidal disruption event black hole mass distribution. This approach, however, has effectively zero constraining power of the value of the low mass slope $\alpha_l$, owing to a lack of sources detected at very low luminosities. The only requirement picked out by the data is $\alpha_l > 0$, so that the TDE rate goes to zero at low masses.  We caution against interpreting the large value of $M_c\sim 10^6 M_\odot$ as a fundamental change in the {\it intrinsic} black hole mass function at this scale, as it is possible that the high value is driven by the lack of observed plateau luminosities below $\log L_{UV, {\rm plat}} \lesssim 40.5$ (see Fig. \ref{fig:thedists}). We cannot distinguish between an intrinsic and volume-limiting effect at this stage, a result we discuss  further in section \ref{sec:implications}. Current surveys do not necessarily have the required observable sensitivity to detect TDEs with plateau luminosities at this low level for the vast majority of our sources, and upper limits were found for some sources in the analysis of  \cite{Mummery_et_al_2024}, which we do not include in this work. We note however that the detection (or not) of a plateau for a given TDE was not found to correlate with galaxy mass \citep[and therefore presumably black hole mass][]{Mummery_et_al_2024}. Future observational surveys with higher sensitivity will likely be able to overcome this effect, and determine if this represents a {\it true} turnover mass (if indeed such a mass scale exists) in the black hole mass function.

This analysis finds $\alpha_h = -0.85 \pm 0.4$ (median $\pm$ standard deviation) and $\log M_c/M_\odot = 5.8 \pm 0.8$. This suggests that the observed TDE population is dominated by black holes with masses $\sim 10^{6.5} M_\odot$.  It is interesting to note that this high mass slope is statistically steeper than expected from extrapolated galactic scaling relationships and conventional models of the observed rate of tidal disruption events on black hole mass \citep[which would predict $\alpha_h \simeq 0$ for Eddington-limited accretion, e.g.,][]{Stone16}.

As a test of this analysis we compare the cumulative distribution function of the model black hole mass distribution with the observed black hole mass distribution, computed in three different manners. These are from the $M_\bullet-M_{\rm gal}$ relationship (green step function, using the relationship of \citealt{Greene20}), the $M_\bullet-\sigma$ relationship (red step function, using the relationship of \citealt{Greene20}), and the $M_\bullet-L_{\rm plat}$ scaling relationship (blue step function, using the relationship of \citealt{Mummery_et_al_2024}). The model  posterior median (purple curve) and $1 \sigma$ uncertainty (purple shaded region), are also shown, with the effects of Hills mass suppression included (the effects of Hills mass suppression are discussed in more detail below). The posterior median model produces a formally acceptable description of both the $M_\bullet-M_{\rm gal}$ and $M_\bullet-L_{\rm plat}$ populations (we cannot reject the null hypothesis that the observed and simulated distributions are the same). The model is in contention with the $M_\bullet-\sigma$ population however ($p$-value = 0.001).

This of course means that different scaling relationships are in contention with one another. Some of this tension is easy to understand physically: neither the $M_\bullet-M_{\rm gal}$ relationship or the $M_\bullet-\sigma$ relationship incorporate the effects of the Hills mass (whereas the plateau luminosity scaling relationship does), and therefore the deviations between the different scaling relationships at $M_\bullet \gtrsim 10^{7.5} M_\odot$ are artificial. The origin of the discrepancy between the $M_\bullet-M_{\rm gal}$ and $M_\bullet-L_{\rm plat}$ relationships and the $M_\bullet-\sigma$ relationship at low black hole masses is less clear. We note that galactic scaling relationships have large intrinsic scatter (typically at least 0.5 dex), and so the smaller the sample (as in the case of $\sigma$ measurements of which we have less than half of the number of $M_{\rm gal}$ measurements) the larger the uncertainty in the cumulative distribution function. Furthermore, the \cite{Greene20} scaling relationships are calibrated on substantially larger black hole masses than the mass region of discrepancy (see e.g., Fig. \ref{fig:new_scaling}), and it is unclear (physically, as well as empirically) whether these relationships potentially flatten at lower masses, which would remove this tension. As TDE sample sizes increase in the Rubin/LSST era this will become an increasingly interesting question. 

With the parameters of the black hole mass distribution constrained, we turn to the ability of this model to reproduce the observed luminosity functions, beginning with the plateau luminosity function.

\subsection{The late-time UV plateau distribution }
\begin{figure}
    \centering
    \includegraphics[width=\linewidth]{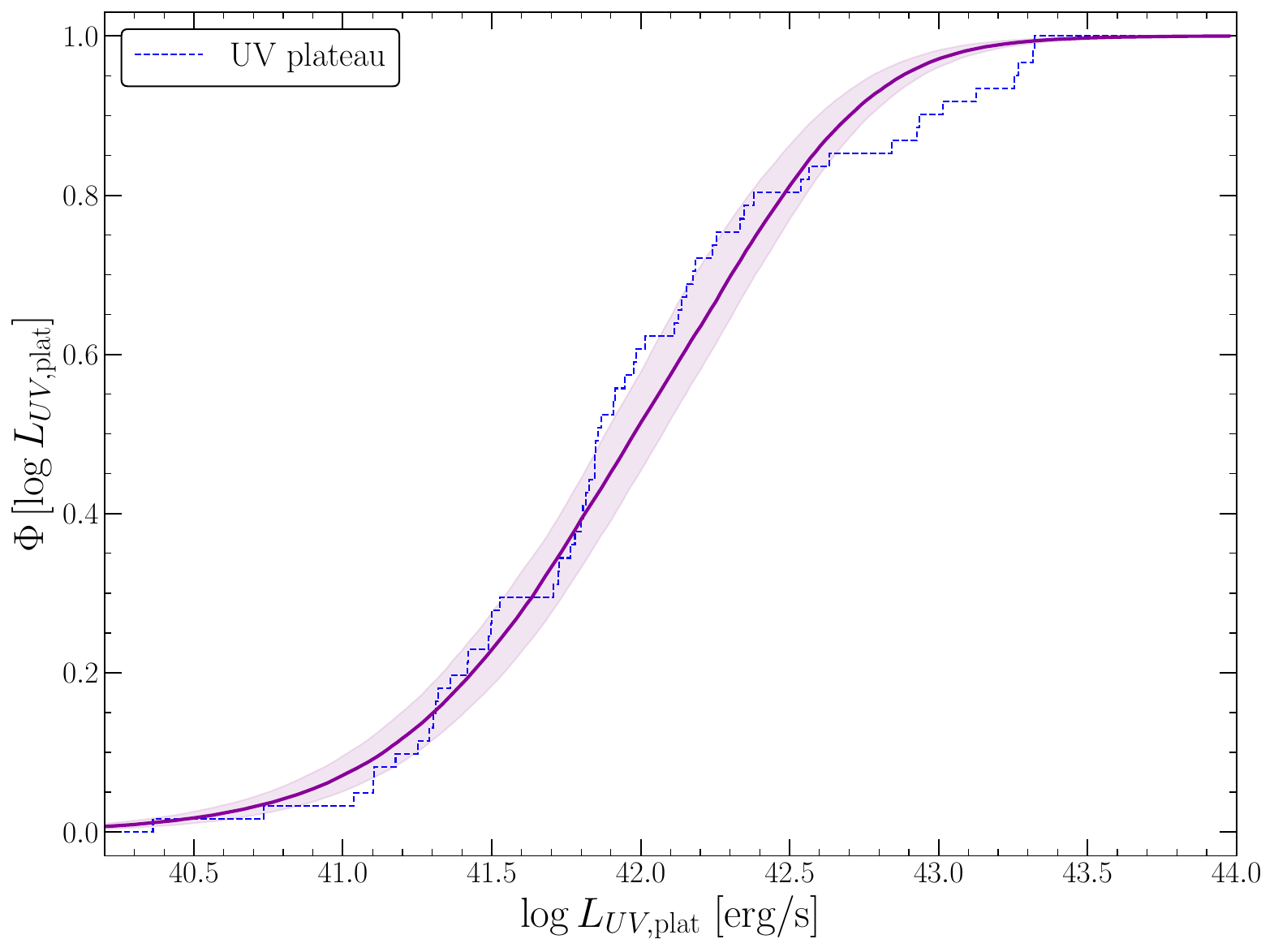}
    \includegraphics[width=\linewidth]{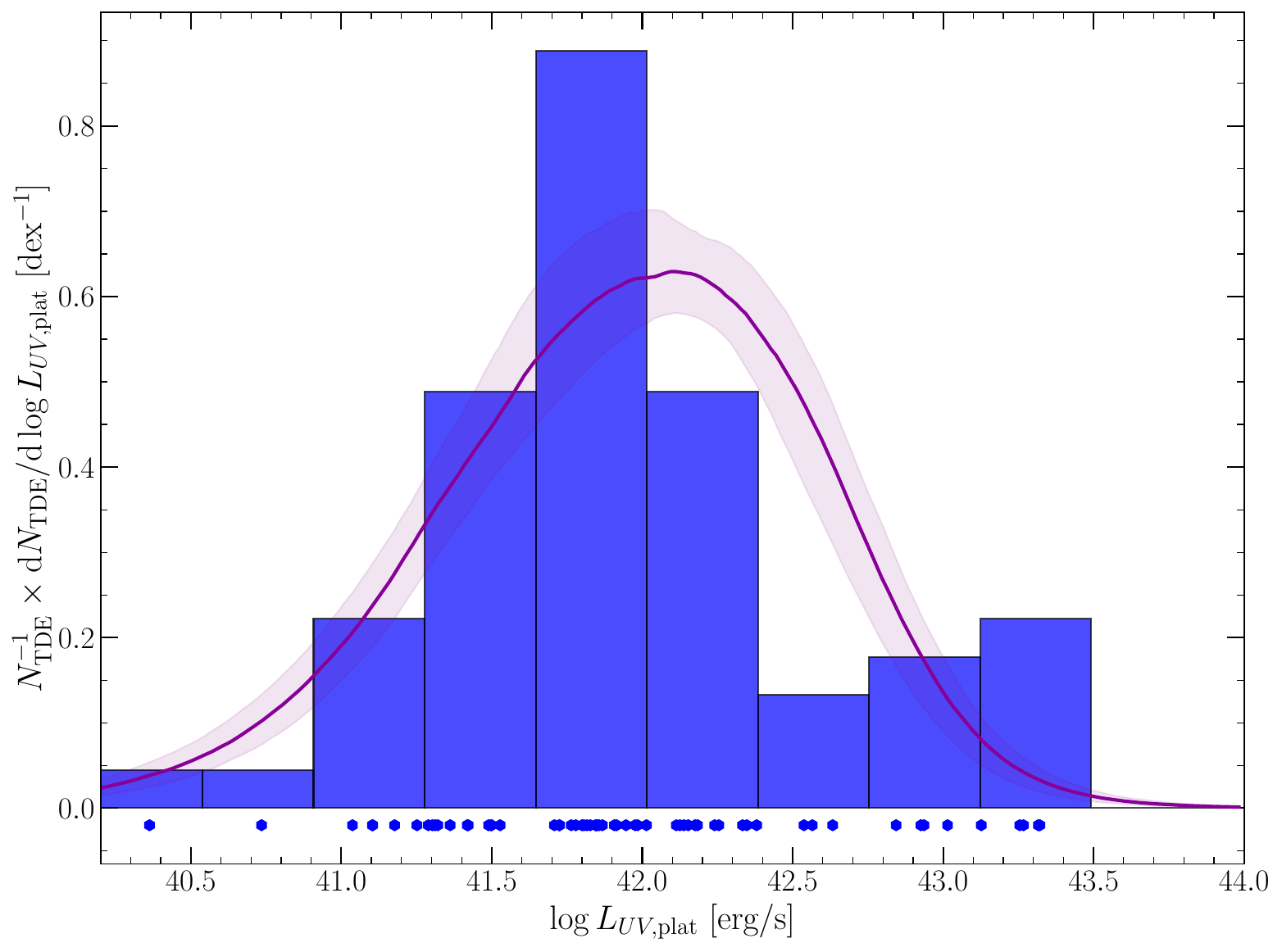}
    \includegraphics[width=\linewidth]{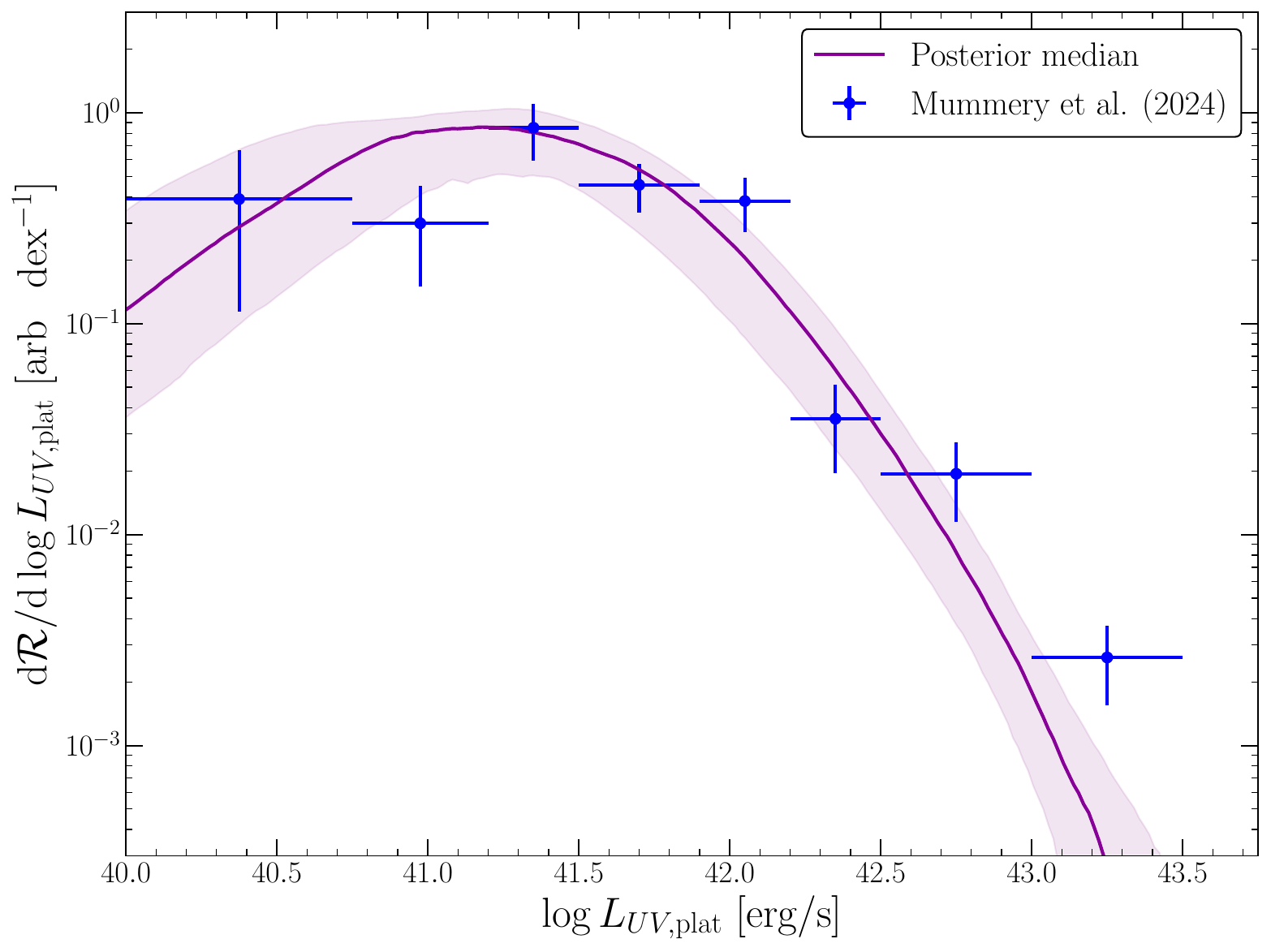}    
    \caption{Lower: The luminosity function of the UV plateau distribution, with each observation weighted by the volume associated with its peak optical luminosity (blue points). Also shown are the model predictions derived in this work (purple solid curve, with $1\sigma$ uncertainty shown by the shaded region).   Middle: the probability density function (unweighted) of the observed UV plateau luminosity sample (blue histogram; raw data shown by points), and posterior median (purple curve) and $1 \sigma$ uncertainty (purple shaded region). Upper: the cumulative distribution function of the observed UV plateau luminosity sample (blue step function) and the model  posterior median (purple curve) and $1 \sigma$ uncertainty (purple shaded region). The model produces an acceptable fit to the observed plateau luminosities.  The null hypothesis that the observed and simulated distributions are the same cannot be rejected by a two sample KS test ($p$-value = 0.30). }
    \label{fig:platfuncs}
\end{figure}

As a first test of the theory, we compare the observed  distributions of the UV plateau luminosity, and the corresponding distributions predicted from the MCMC posterior median. Our results are displayed in Figure \ref{fig:platfuncs}.

The middle panel of Figure \ref{fig:platfuncs} shows the observed plateau luminosity distribution in our sample (blue histogram and points), with model posterior shown by the purple curve and shaded region (median and $1\sigma$ uncertainty). To construct these curves we sample $N = 10^5$ tidal disruption events following the procedure described above, with $\alpha_l, \alpha_h$ and $M_c$ set by their posterior median values (or sampled from the posterior distribution for each curve in the computation of the shaded region).  The distribution of these $10^5$ luminosities is then the predicted UV plateau distribution.  As can be seen in the middle and upper panels of Fig. \ref{fig:platfuncs}, the fit is acceptable. 

In the lowest panel in Figure \ref{fig:platfuncs} we also display the (unnormalised) luminosity function of the UV plateau population. The data is as in Figure \ref{fig:thedists}, and discussed above. The only slight change required to the theoretical distribution is to correct for $1/{\cal V}_{\rm max}$ in an analytical manner, which we do using the empirical peak optical luminosity-black hole mass scaling derived in \cite{Mummery_et_al_2024} (discussed in more detail later, see equation \ref{eq:lpeak}).  In practice this corresponds to merely modifying each $\alpha_l, \alpha_h$ when sampled by $\alpha_l, \alpha_h \to \alpha_l -1.53, \alpha_h-1.53$. This distribution provides an acceptable description of the observed UV plateau luminosity function.

The fit to the probability density function and cumulative distribution function of the UV plateau data is formally acceptable. The null hypothesis that the observed and simulated distributions are the same cannot be rejected by a two sample KS test ($p$-value = 0.30). This result gives us confidence in our disc theory approach, but the most important parts of this analysis relate to the implications of these results {\it which are not part of the fit}, e.g., the ability of this model to reproduce the X-ray luminosity distribution. 

\begin{figure}
\centering 
    \includegraphics[width=\linewidth]{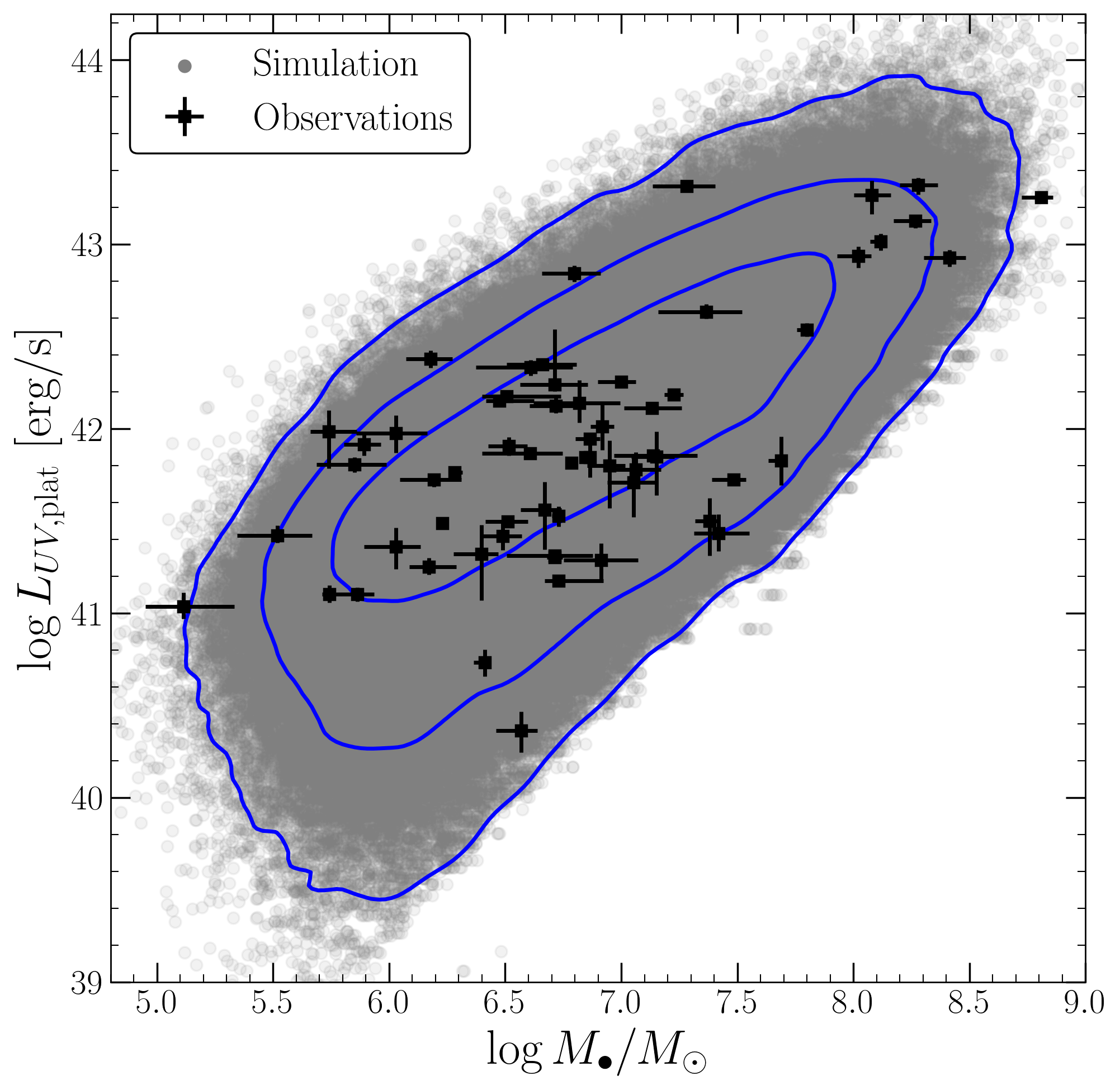}
    \caption{The two-dimensional mass-plateau luminosity plane of the model derived here and observational data. In grey we plot a sample of $10^6$ theoretical TDEs, drawn from the posterior median black hole mass distribution, with the observational TDE data overplotted. Note that the intrinsic scatter in black hole measurements are not shown, and for the galactic mass scaling relationship this is of order 0.8 dex \citealt{Greene20}.  Blue contours show 1, 2 and 3 sigma, i.e., they contain the central $68.2\%, 95.4\%$ and $99.7\%$ of the simulated data respectively.  }
    \label{fig:platsamp}
\end{figure}

An example of one of these non-fitted results is shown in Figure \ref{fig:platsamp}. In Figure \ref{fig:platsamp} we display the two dimensional black hole mass-plateau luminosity plane of our posterior median results. This was formed from a sample of $N = 10^6$ TDE systems following the procedure set out above. Each individual TDE sample is displayed by a grey point, with the blue curves showing  1, 2 and 3 sigma contours (i.e., they contain the central  $68.2\%, 95.4\%$ and $99.7\%$ of the simulated data respectively). Also displayed are the observed TDE plateau luminosities, with black hole masses inferred from the galactic mass scaling relationship (we choose this rather than the velocity dispersion as not every TDE has a published velocity dispersion measurement). Note that the intrinsic scatter in each TDEs black hole mass measurement are not shown, and for the galactic mass scaling relationship this is of order 0.8 dex \citep{Greene20}. 

It is clear that the model put forward in this paper can reproduce not only the one-dimensional distribution of the TDE plateau luminosity population (Fig. \ref{fig:platfuncs}), but also the more detailed two-dimensional mass-plateau luminosity plane (Fig. \ref{fig:platsamp}). 

\subsection{The peak X-ray luminosity distribution}

\begin{figure}
    \centering
    \includegraphics[width=\linewidth]{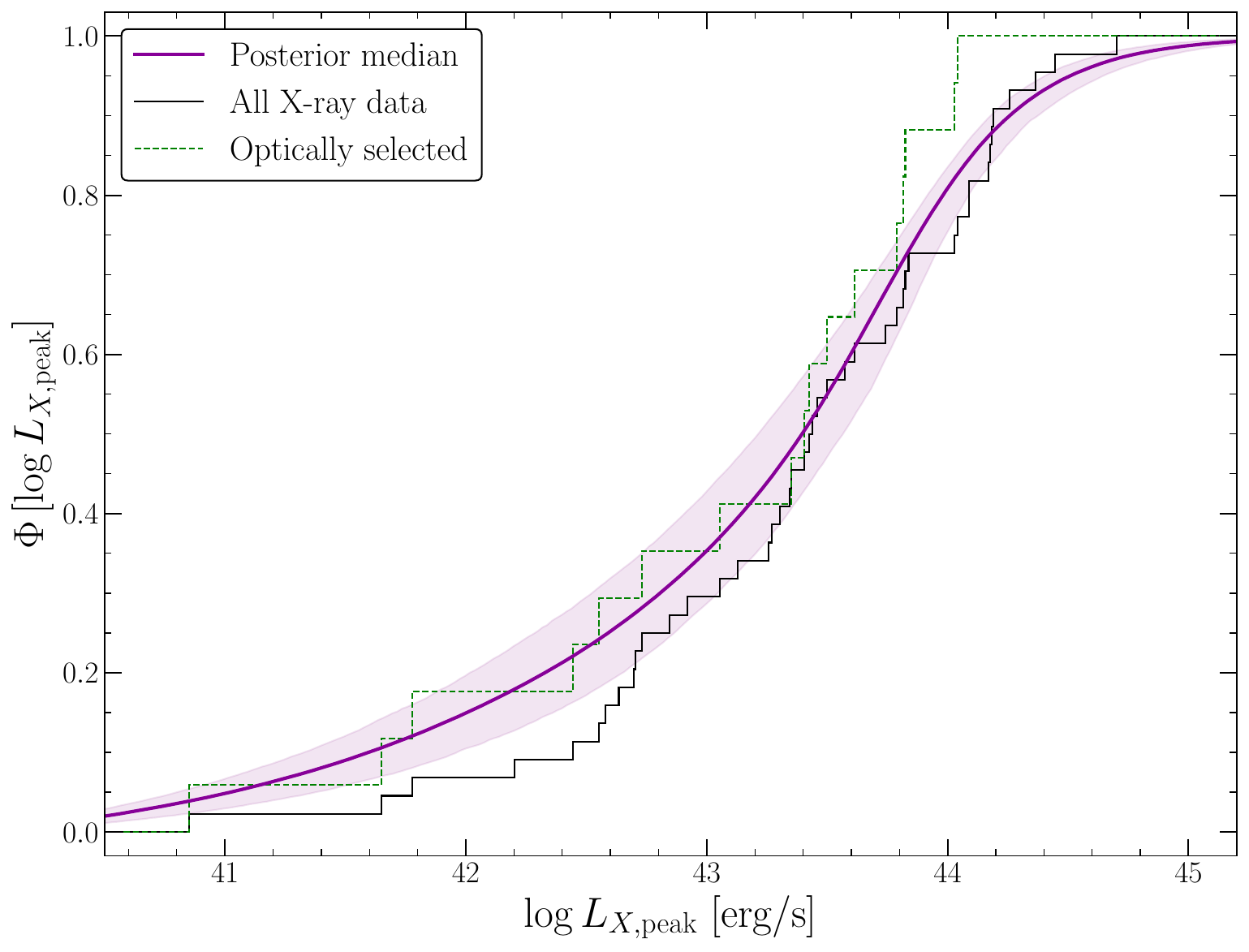}
    \includegraphics[width=\linewidth]{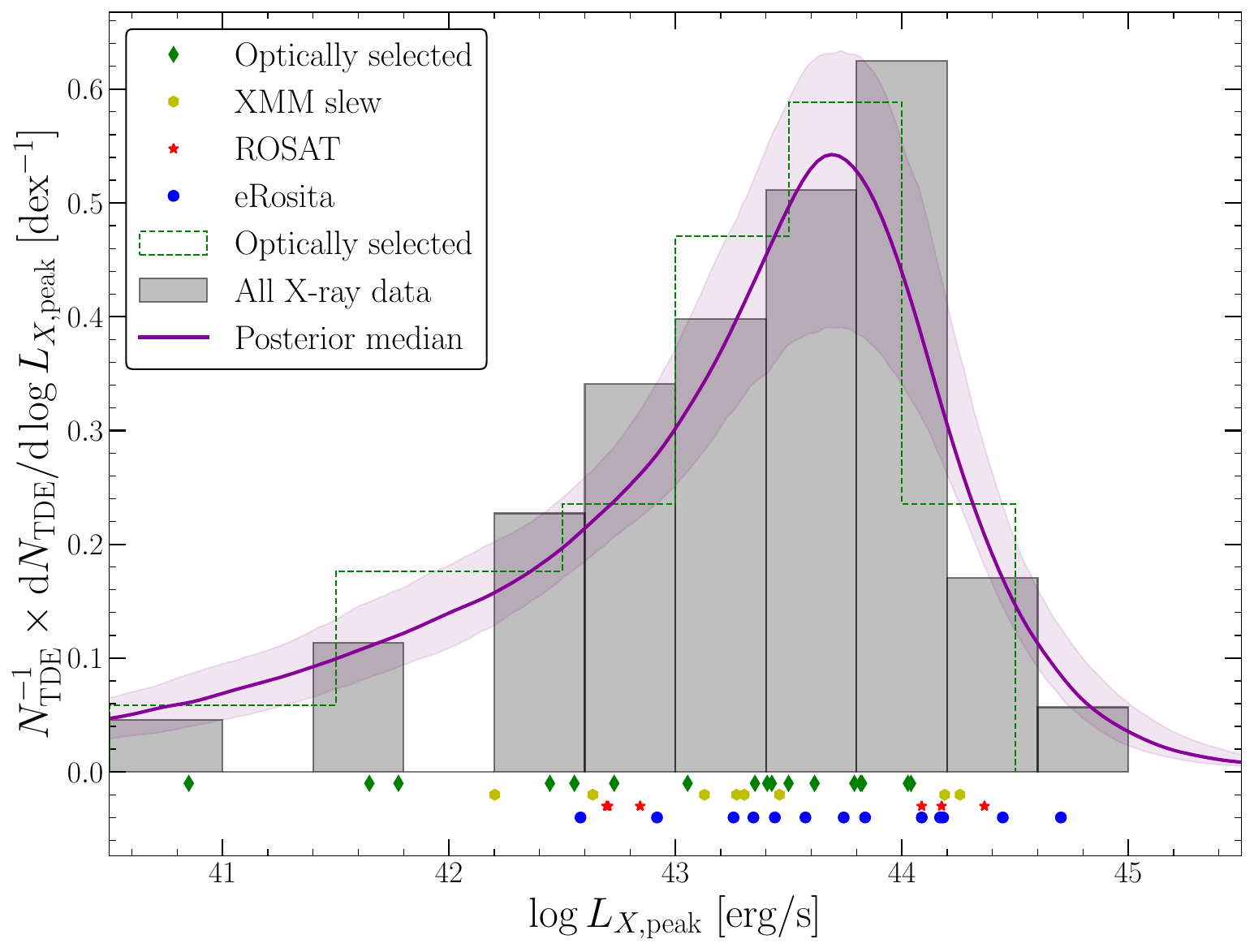}
    \includegraphics[width=\linewidth]{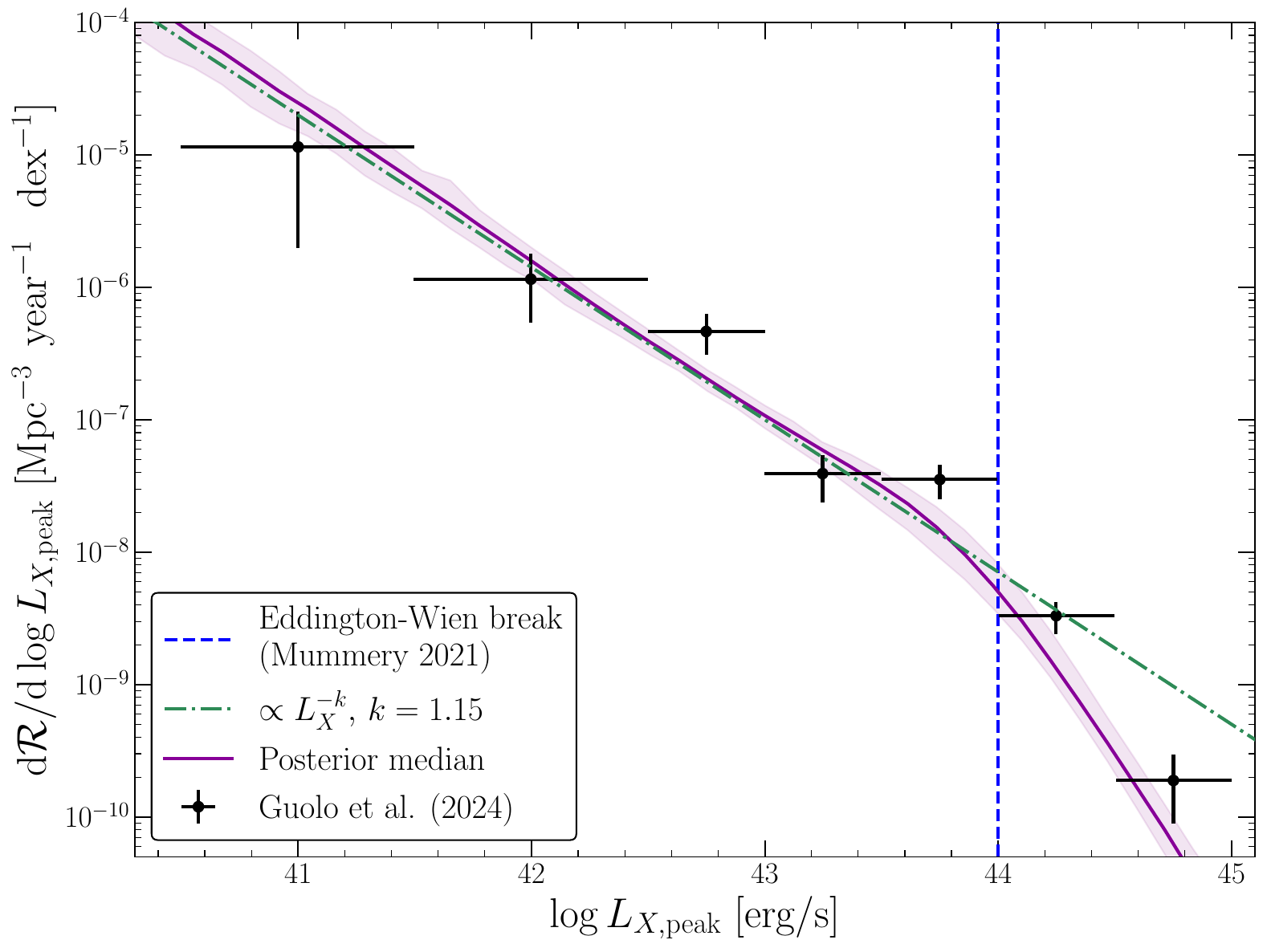}
    \caption{Lower: the inferred rate of the X-ray peak luminosities of the tidal disruption event population (black points), and the distribution predicted by the posterior fit to the UV plateau data (purple line and $1\sigma$ shaded uncertainties). Middle: the probability density function (unweighted) of the observed peak X-ray luminosity sample (black histogram; raw data shown by individual points) coloured by detection method), and posterior median (purple curve and $1\sigma$ shaded uncertainties). Upper: the cumulative distribution function of the observed peak X-ray luminosity sample (black step function) and the model posterior (purple curve and $1\sigma$ shaded uncertainties). The model produces an excellent fit to all of the statistical properties of the data.  The null hypothesis that the observed and simulated distributions are the same cannot be rejected by a two sample KS test ($p$-value = 0.41).  }
    \label{fig:xrayfuncs}
\end{figure}

The ability of a relativistic accretion disc model, coupled to a parameterised black hole mass distribution, to reproduce the observed tidal disruption event UV plateau distribution is a promising result. More important however is the question of whether this same model can reproduce other, independent,  observed properties of the tidal disruption event distributions, which are also expected to result from direct emission from accretion flows. This test is important because it would demonstrate that the physics of the model is correct, and the parameters of the black hole mass distribution which we obtained from a fit to the late-time UV luminosity distribution are in fact meaningful, and are not simply being overfit to the data. 

In addition to the late-time optical/UV luminosity of tidal disruption event systems, the X-ray luminosity of these systems is also thought to be powered by accretion, a framework successfully reproduces  the data of numerous individual sources \citep{MumBalb20a, Wen20, Wen21, Mummery_Wevers_23, Guolo24}.  If both the model and statistical framework put forward in this work are correct, then this approach should naturally reproduce the X-ray luminosity function constructed in \cite{Guolo24} with no further free parameters.   

To perform this test we numerically sample a large $N = 10^5$ sample of tidal disruption event X-ray luminosities (defined in this work as the integrated $0.3-10$ keV disc luminosity). We use the median probability density function for the black hole mass distribution $(N_{\rm TDE}^{-1}  {{\rm d} N_{\rm TDE} / {\rm d} \log M_\bullet})$ obtained for the UV plateau distribution, and resample  sources following an identical procedure as spelled out in \cite{Mummery_et_al_2024} and Appendix \ref{app:stats}. We then further sample from the posterior distribution of $\alpha_l, \alpha_h$ and $M_c$, again sampling $10^5$ X-ray TDEs each time, to construct a theoretical uncertainty in the X-ray luminosity distribution.  For each sampled TDE we then determine the peak X-ray luminosity (which is the most natural quantity to examine on the population level, and who's luminosity function was constructed by \citealt{Guolo24}). All of the data used in this section is taken from \cite{Guolo24}, including the raw luminosities, the source designation (e.g., optically selected versus XMM slew), and X-ray  luminosity function. 

Some care must be taken at this point in this analysis, as at early times in the innermost disc regions relevant for X-ray observations (as opposed to the much later times and larger radii relevant for the UV plateau), some of the sampled systems may become formally super-Eddington. At super-Eddington luminosities the governing assumptions of the relativistic accretion theory employed in this work begin to break down, as the flow enters a so-called ``slim'' disc state \citep[][see \citealt{Wen20} for an application of this model to tidal disruption events]{Abram88}.  No out of temporal-equilibrium relativistic slim disc theory currently exists, as only steady state theories have been developed. While a steady state theory is unlikely to strictly correctly describe the properties of a tidal disruption event accretion flow near peak luminosity, as this is the evolutionary point at which  the steady state assumptions are maximally violated, steady state theory does highlight a key piece of physics which can be incorporated into the thin disc theory used in this work. In a steady state theory an accretion flow is parameterised by a constant radial mass flux $\dot m$. For  sub-Eddington accretion the disc bolometric luminosity (normalised by the Eddington luminosity and denoted $l$) scales linearly with this accretion rate $l \propto \dot m$.  Above the Eddington accretion rate however, an increase in $\dot m$ only results in a logarithmic increase in the emitted luminosity \citep{SS73}, namely $l \propto \dot m_{\rm edd} (1 + \ln \dot m)$. The physical reason for this transition from a linear to logarithmic increase is that the extra liberated energy above and beyond the Eddington limit is nearly all advected with the flow, rather than radiated away (energy advection is negligable in thin disc systems).   As any apparent super-Eddington emission will in all likelihood  be advected with the flow, it is a good approximation to treat the maximum bolometric luminosity of these sources as being fixed at the Eddington luminosity. We therefore compute the bolometric disc luminosity of each of our sampled systems, in addition to the $0.3-10$ keV X-ray luminosity.  If the bolometric luminosity ever exceeds the Eddington, then the disc evolution is stopped (at a time we denote $t_{\rm edd}$), and the peak X-ray luminosity of this system is recorded at this time $L_{X, {\rm peak}} = L_X(t_{\rm edd})$. Otherwise, the peak X-ray luminosity is recorded as whatever maximum value the disc reaches during its evolution (this is typically at a time roughly one viscous timescale into the evolution of the flow). This Eddington-limited framework was first proposed for tidal disruption event systems by \cite{Mum21}.

\begin{figure}
\centering 
    \includegraphics[width=\linewidth]{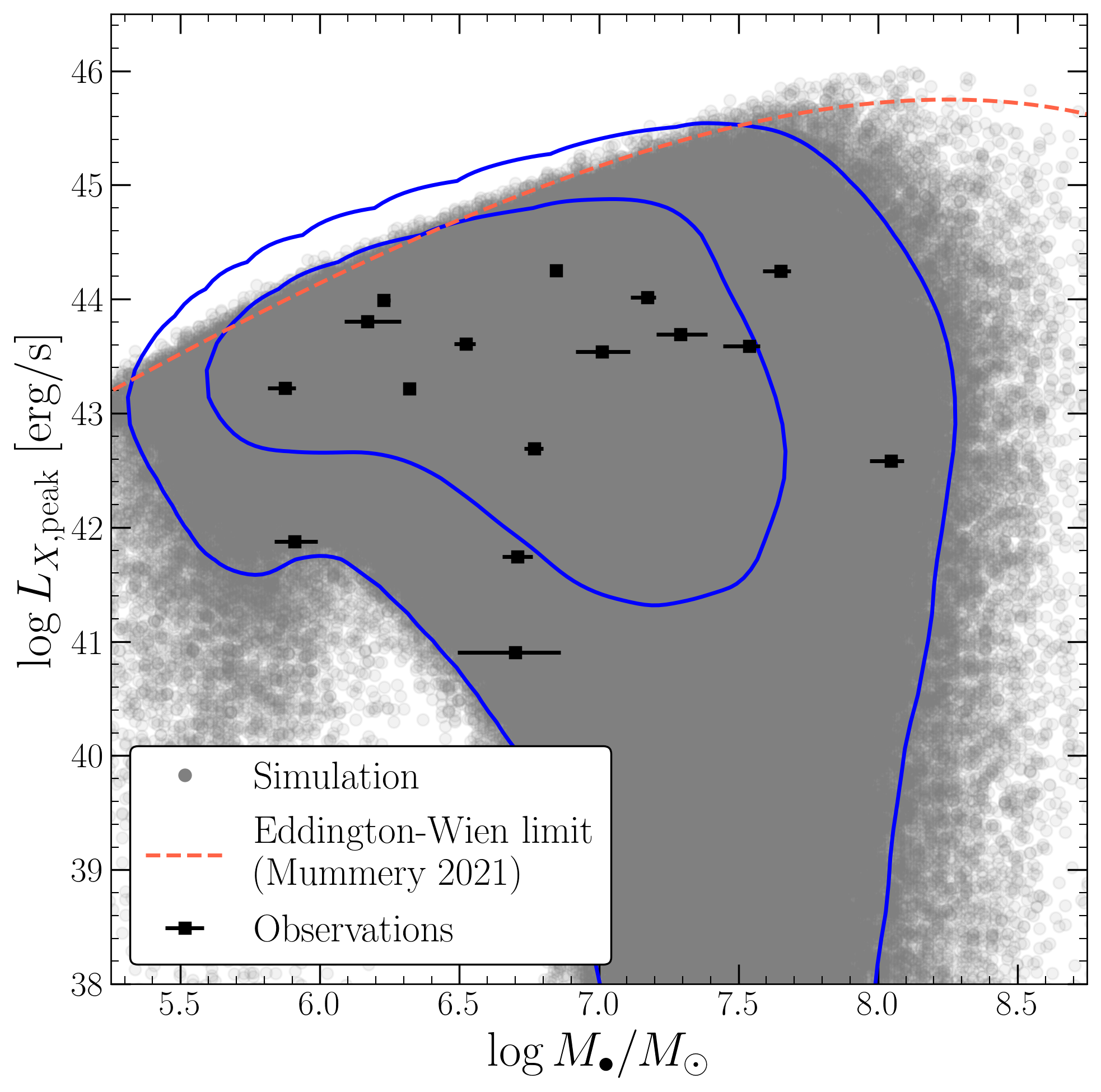}
    \caption{The two-dimensional mass-peak X-ray luminosity plane of the model derived here and observational data. In grey we plot a sample of $10^6$ theoretical TDEs, drawn from the posterior median black hole mass distribution, with the observational TDE data overplotted. Note that the intrinsic scatter in black hole measurements are not shown, and for the galactic mass scaling relationship this is of order 0.8 dex \citealt{Greene20}.  Blue contours show 1 and 2 sigma, i.e., they contain the central $68.2\%$ and $95.4\%$ of the simulated data respectively. The red dashed curve shows the Eddington-Wien limiting luminosity derived in \citealt{Mum21}.  }
    \label{fig:xraysamp}
\end{figure}

\begin{figure}
\centering 
    \includegraphics[width=\linewidth]{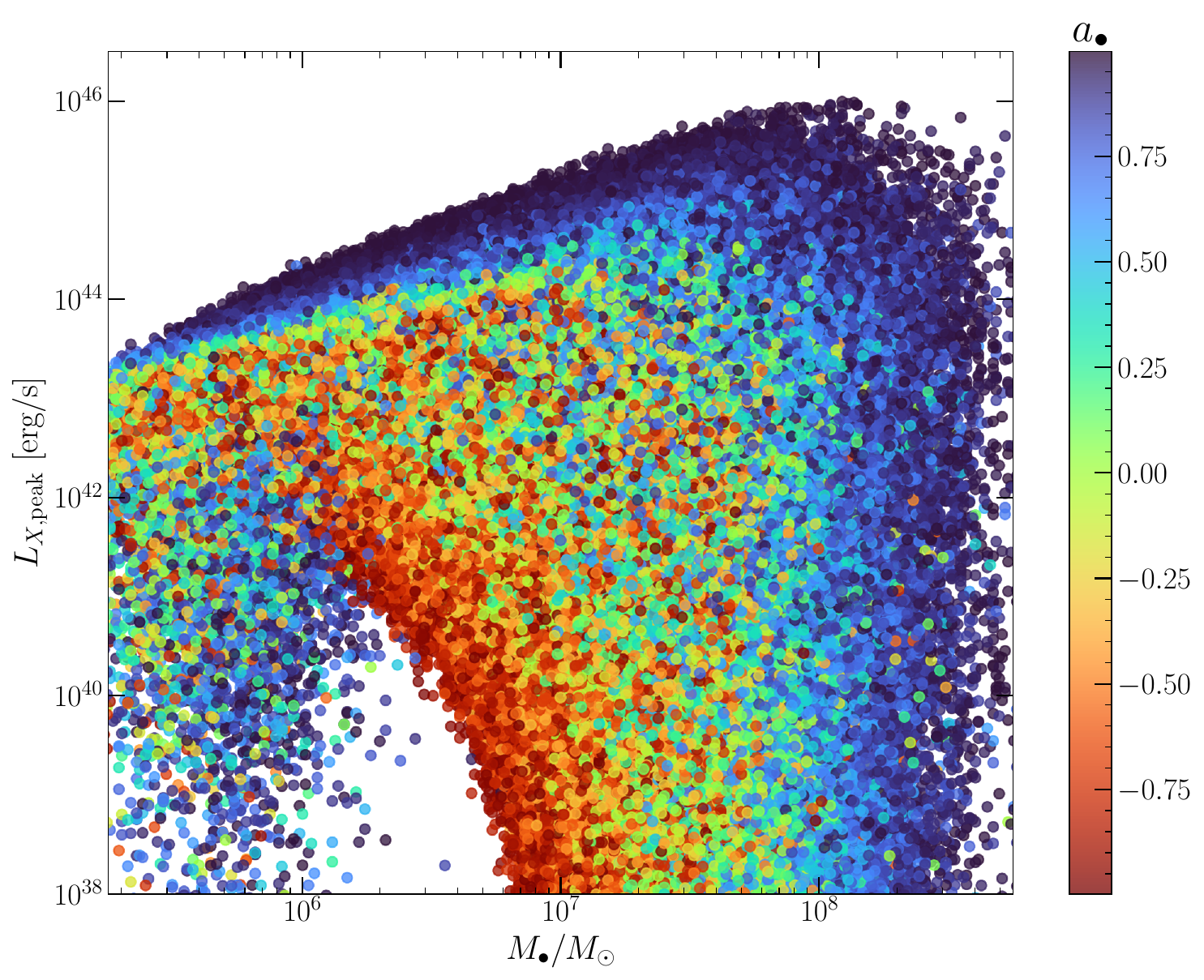}
    \caption{The three-dimensional mass-peak X-ray luminosity-spin plane of the model derived here. By points we display a sample of $N=10^6$ theoretical TDEs, drawn from the posterior median black hole mass distribution, with each point coloured by the spin parameter of the central black hole. Note that the relative density of points is much more difficult to ascertain in this figure, but can be identified with reference to Fig. \ref{fig:xraysamp}. There are two key trends with black hole spin which can be seen here, the first is the increasing spin-dependent Hills mass allowing bright X-ray emission from more massive black holes (the left-to-right trend), while more rapidly spinning black holes produce brighter X-ray emission at all masses (the trend bottom-to-top).  }
    \label{fig:xrayspin}
\end{figure}

A key prediction made in \cite{Mum21}, was that this Eddington limiting (coupled with the emission of TDEs typically being in the Wien tail), would lead to an observable change in the X-ray luminosity distribution at the characteristic ``Eddington-Wien'' luminosity scale $L_{X -\rm EW} \simeq 10^{44}$ erg/s, with significantly fewer sources detected with luminosities above this cutoff. 

The results of this analysis, having followed the  sampling procedure discussed above, are displayed in Figure \ref{fig:xrayfuncs}.  To go from the sampled X-ray luminosity distribution to the intrinsic X-ray luminosity function it suffices to weight each system in our sample by $L_{X, {\rm peak}}^{-3/2}$ (to compensate for the Malmquist bias experienced by low luminosity sources). The normalisation of this rate is arbitrary (in our model), and was fit to match the amplitude of the \cite{Guolo24} analysis. 

As is clear from Figure \ref{fig:xrayfuncs}, relativistic disc theory is capable of reproducing the tidal disruption event X-ray luminosity function, from an original fit to the late time UV luminosity function, with no additional free parameters.  The null hypothesis that the observed and simulated distributions are the same cannot be rejected by a two sample KS test ($p$-value = 0.41).  This is an extremely important result, as it demonstrates that the accretion flows which produce the late-time plateau observed in optical/UV bands are the very same accretion flows which produce the early time X-ray luminosity observed in many systems. 

The relevance of the Eddington-Wien upper luminosity limit can be clearly seen in both the probability density function of the peak X-ray luminosity population, and the TDE X-ray luminosity rate. In the raw observed distribution there is a clear turnover in the observed luminosity distribution at $L_{X \rm -EW} \simeq 10^{44}$ erg/s, while the peak X-ray luminosity rate displays a clear break at the same luminosity scale \citep[as was first described empiraclly by][]{Guolo24}. Below the break the peak X-ray luminosity rate (of both the data and model) can be well described by a power-law of index $k \simeq -1.15$

Unlike the earlier UV plateau analysis, not every tidal disruption event in the \cite{Guolo24} sample comes from an optical all-sky survey. In addition, not  every observed tidal disruption event is comprehensively followed up with X-ray observations, and so it is natural to worry somewhat about combining different samples in this way.  To examine if this survey combining has any effect on the strength of our results we can cary out an identical analysis using only those tidal disruption events discovered in optical surveys (the green dashed curves in Figure \ref{fig:xrayfuncs}).   The null hypothesis that the observed and simulated distributions are the same in this case also cannot be rejected (a two sample KS test returns a $p$-value = 0.62).   

Similarly, an identical analysis of those tidal disruption events discovered only by X-ray surveys yields similar results. The null hypothesis that the observed and simulated distributions are the same in this case also cannot be rejected (a two sample KS test returns a $p$-value = 0.14).   

This strength of this result can again be further emphasised by examining the two-dimensional mass-luminosity plane of the theory and data. In Figure \ref{fig:xraysamp} we display the two dimensional black hole mass-peak X-ray luminosity plane of our posterior median results. This was formed from a sample of $N = 10^6$ TDE systems following the procedure set out above. Each individual TDE sample is displayed by a grey point, with the blue curves showing  1 and 2 sigma contours (i.e., they contain the central  $68.2\%$ and $95.4\%$ of the simulated data respectively). Also displayed are the observed optically-selected TDE peak X-ray luminosities, with black hole masses inferred from the galactic mass scaling relationship (we choose this rather than the velocity dispersion as not every TDE has a published velocity dispersion measurement). Note that the intrinsic scatter in each TDEs black hole mass measurement are not shown, and for the galactic mass scaling relationship this is of order 0.8 dex \citep{Greene20}. 

These results strongly suggest that the optically and X-ray selected population of tidal disruption events are not distinct, but are drawn from the same overall distribution.

In addition, in Figure \ref{fig:xrayspin} we display an interesting physical property of the X-ray luminosity of the tidal disruption event population. Figure \ref{fig:xrayspin} shows the three-dimensional mass-peak X-ray luminosity-spin plane of the model derived here. By points we display a sample of $N=10^6$ theoretical TDEs (the same sample as in Fig. \ref{fig:xraysamp}), drawn from the posterior median black hole mass distribution, with each point coloured by the spin parameter of the central black hole. Note that the relative density of points is much more difficult to ascertain in this figure, but can be identified with reference to Fig. \ref{fig:xraysamp}. There are two key trends with black hole spin which can be seen here, the first is the increasing spin-dependent Hills mass allowing bright X-ray emission from more massive black holes (the left-to-right trend), while more rapidly spinning black holes produce brighter X-ray emission at all masses (the trend bottom-to-top). This contrasts interestingly with the UV plateau luminosity \citep[see Fig. 5 of][]{Mummery_et_al_2024}, which is only sensitively to the Hills mass effect, not the intrinsic luminosity effect. Physically this results from the sensitivity of the X-ray luminosity in a TDE system to the temperature of the inner disc; this inner disc temperature is sensitive to the black hole spin parameter (more rapidly rotating black holes have a higher accretion efficiency $\eta$, and liberate more of the available accretion energy as photons).

Finally, we discuss the implications for our X-ray luminosity result in the context of different ``unification'' schemes for understanding TDE emission \citep[e.g., the reprocessing model of][]{Dai18}. We note that we are able to reproduce the X-ray luminosity function, and observed distribution, of TDEs without requiring any absorption of X-ray photons along the line of sight between the disc and the observer. We also note that the optically-selected and X-ray-discovered TDEs show near identical X-ray luminosity distributions (Fig. \ref{fig:xrayfuncs}). At first glance this seems to suggest that the early time optical luminosity is not produced by the reprocessing (and therefore absorption) of X-ray photons from an accretion flow. However, we caution that there are other interpretations of this result.  We cannot rule out that some fraction of TDEs yield X-ray non-detections purely as a result of  absorption/reprocessing of X-ray photons,  and we therefore reproduce the observed luminosity function as it contains {\it only} those sources which are not absorbed at all. In fact, roughly 3/4 of our TDE sample (i.e., 75\% of the systems sampled in producing Figure \ref{fig:xraysamp}) are predicted to be X-ray loud ($L_X\gtrsim 10^{42}$ erg/s), while only $\sim 40\%$ of observed optically-selected TDEs satisfy this criterion \cite{Guolo24}, although this is of course a lower-limit owing to incomplete X-ray follow up of optical TDEs.  If this difference in X-ray brightness is in fact related to intrinsic physics, rather than observational selection effects, then we can envision three plausible interpretations of this result.

Firstly, the reprocessing/absorption of X-ray photons (or whatever suppresses the X-ray luminosity in some cases) does occur for some TDE sources, but acts in an all-or-nothing fashion, either completely suppressing the X-ray luminosity of a TDE, or not at all. The reason this interpretation is permitted is that a more subtle change in X-ray luminosity of sources by reprocessing would alter the shape of the luminosity function itself (as opposed to simply reducing the amplitude, the result of removing some sources entirely). We believe this is the least likely of the three interpretations.  

Alternatively, reprocessing is an important process for powering the early time optical emission, but this reprocessing only suppresses the observed X-ray luminosity for a timescale $t_{\rm sup}$ which is {\it shorter than} the typical timescale over which the peak X-ray luminosity of the disc varies significantly. As the X-ray variation timescale is of order a few ``viscous'' timescales, which can comfortably be of order ${\cal O}(100$'s) of days, this is certainly possible. This interpretation is supported by the findings of \cite{Guolo24}, which show a number of sources which appear to be dimmer at early times than expected from their emission observed at late times. 

A third interpretation is that reprocessing of X-ray photons is not the physical mechanism by which the early time optical emission is powered, and the difference between the 3/4 of sources predicted to be X-ray loud and the observed rate of $\sim 40\%$ (if not selection effects) is due to some fraction of TDE sources failing to efficiently form an accretion flow in the earliest stages of their evolution. This may also explain the roughly $\sim 1/3$ of TDE systems which are (currently) not observed to transition to a plateau state in the optical/UV. 

One way to test this going forward would be to examine TDEs with black hole masses $M_\bullet \lesssim 10^{6.0}M_\odot$, with robust detections of discs at late times, which our model predicts would nearly always produce $L_X \gtrsim 10^{42}$ erg/s in an un-absorbed scenario (Fig. \ref{fig:xraysamp}). If a black hole mass could be reliably inferred at this scale while a robust upper limit on the X-ray luminosity was found, this would imply either substantial absorption of X-ray photons or inefficient disc formation. \cite{Guolo24} found that the black hole mass distribution of TDEs with $L_X \geq 10^{42}$ erg/s and $L_X \leq 10^{42}$ erg/s are statistically indistinguishable, with both populations stretching down to $M_\bullet \sim 10^5M_\odot$. On the other hand our results (in the un-absorbed limit) would suggest that X-ray quiet ($L_X \leq 10^{42}$ erg/s) TDEs should be found around {\it slightly} higher black hole masses, and should certainly not extend down to $\sim 10^5 M_\odot$.  While the results of \cite{Guolo24} are based on a relatively small number ($N = 15$) of galactic scaling relationship black hole masses (which have substantial intrinsic scatter), this is suggestive that some TDEs are not producing any observable X-ray emission for some reason which is not related to the accretion flow itself. This result will be tested further as more TDEs (both X-ray bright and dim) are discovered in the future. 

\subsection{The early-time optical peak distribution}

\begin{figure}
    \centering
    \includegraphics[width=\linewidth]{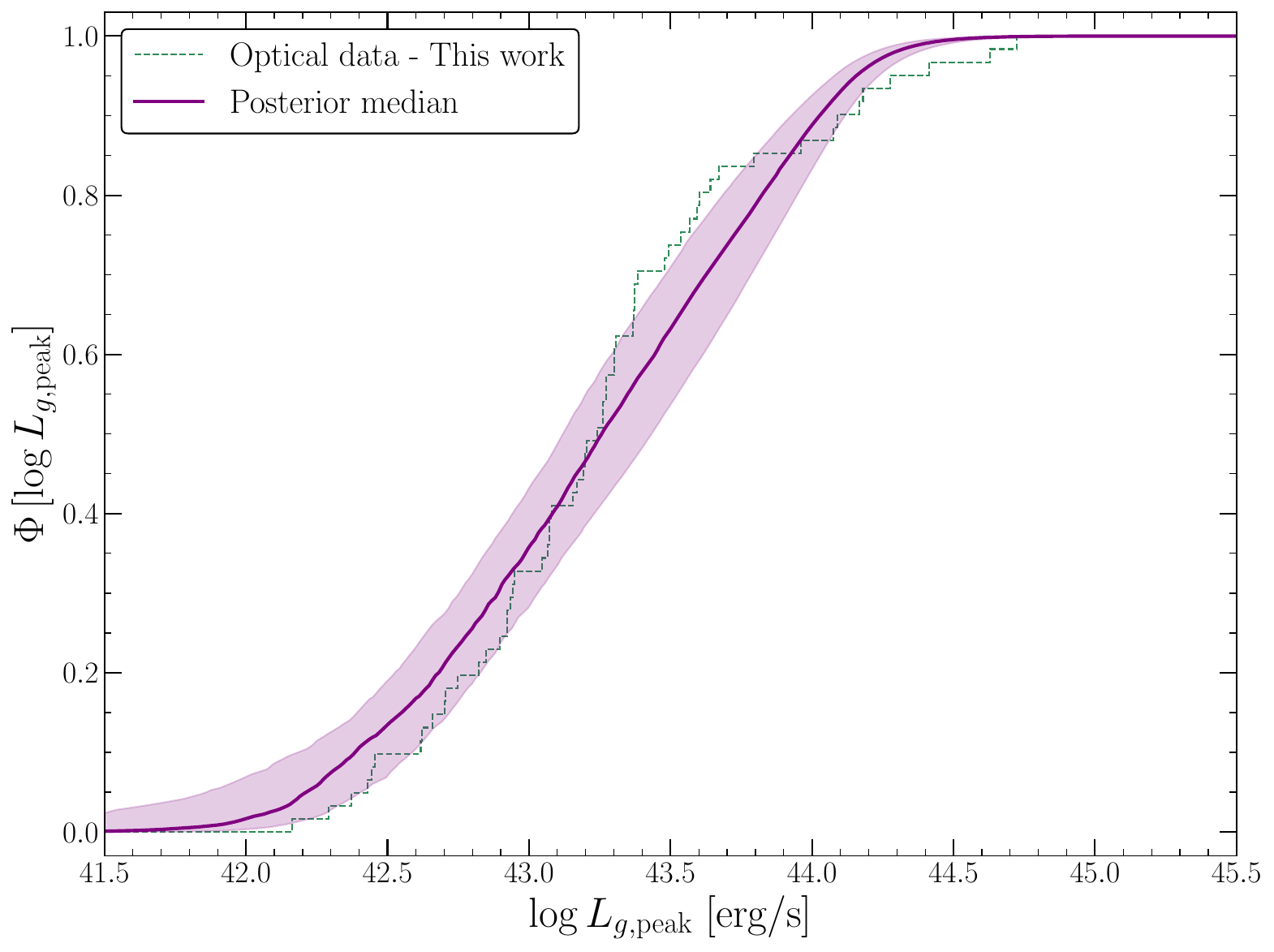}
    \includegraphics[width=\linewidth]{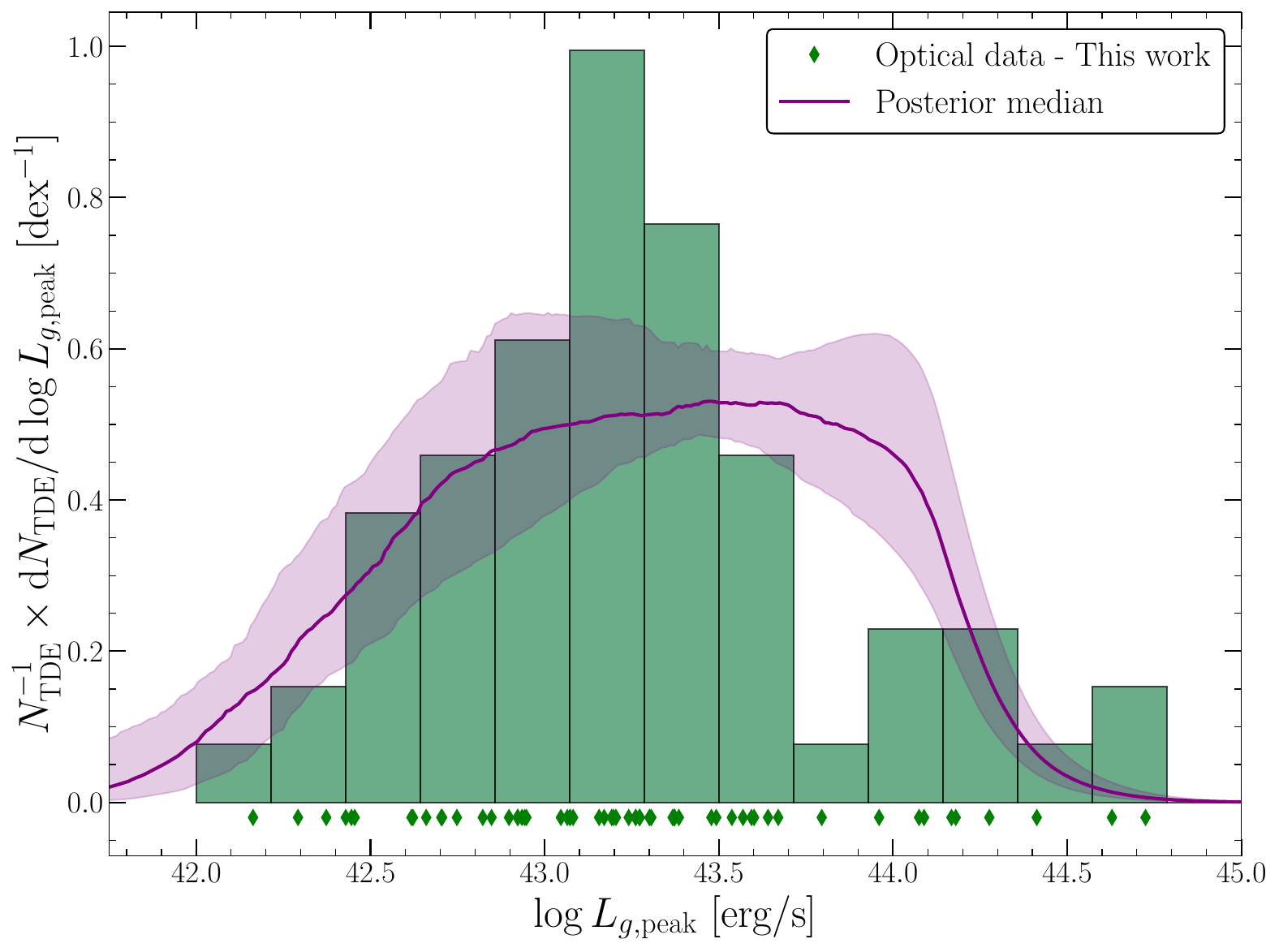}
    \includegraphics[width=\linewidth]{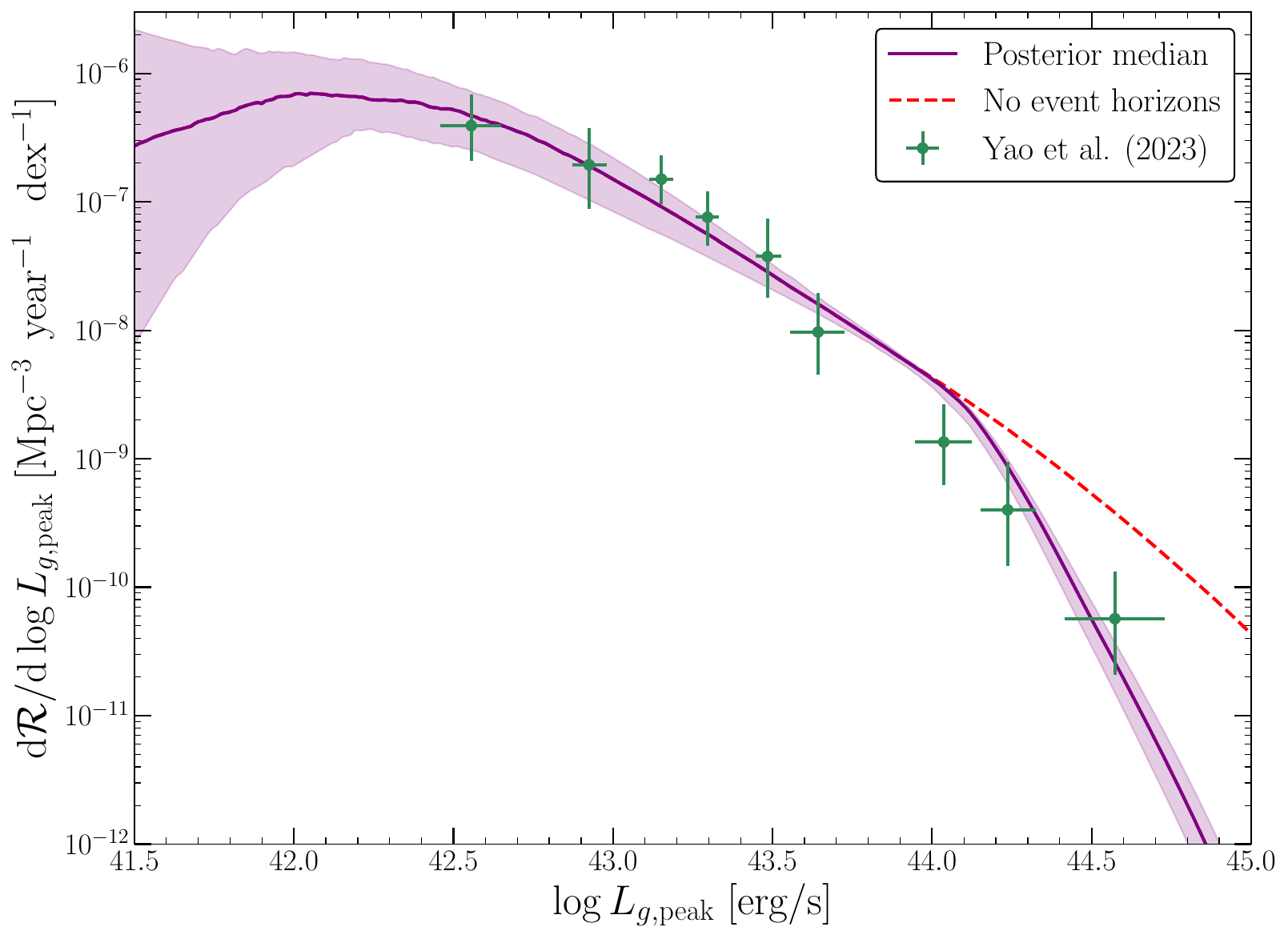}
    \caption{Lower: the inferred rate of the $g$-band peak luminosities of the tidal disruption event population (green points), and the distribution predicted by the posterior median fit the UV plateau data (purple line). Middle: the probability density function (unweighted) of the observed peak $g$-band luminosity sample (green histogram; raw data shown by individual points), and posterior median (purple curve). Upper: the cumulative distribution function of the observed peak $g$-band luminosity sample (green step function) and the model posterior (purple curve). The model produces an excellent fit to all statistical properties of the data.  The null hypothesis that the observed and simulated distributions are the same cannot be rejected by a two sample KS test $p$-value = 0.14. In the lower plot we show the effects of the Hills mass suppression on the observed $g$-band luminosity rate, with the red dashed curve showing the luminosity function if the presence of event horizons is neglected. One sigma model uncertainties in each plot are shown by a purple shaded region.  }
    \label{fig:peakfuncs}
\end{figure}

While the physical origin of the X-ray and late time UV emission in tidal disruption events are clear, the same is not true for the early time optical emission observed in nearly all tidal disruption events, which remains poorly understand. Current models either posit that the observed optical emission results from reprocessed accretion luminosity by a scattering atmosphere \citep[e.g.][]{Guillochon13, Dai18}, from the contraction of a radiation pressure dominated cooling envelope \citep{Metzger22}, or from photon fluxes produced during debris streams shocking during the disc circularisation process \citep[e.g.][]{Piran15, Ryu20}.  As there is no consensus on the physical origin of the early time optical emission, we turn to the empirical scaling relationships derived using the results of disc theory first presented in \cite{Mummery_et_al_2024}.  The raw luminosity data used in this section was taken from \cite{Mummery_et_al_2024}, while the $g$-band luminosity function was taken from \cite{Yao23}. 

It is well known that the early time optical luminosity in a tidal disruption event does not originate from the direct observation of an accretion flow.  However,  disc theory allows the black hole masses in a given disruption to be measured from the late time UV plateau luminosity. Using these masses, \cite{Mummery_et_al_2024} found an empirical relationship between peak $g$-band luminosity and black hole mass. This relationship is given by  
\begin{equation}\label{eq:lpeak}
\log L_{g, {\rm peak}}/{\rm erg}\,{\rm  s}^{-1} = 1.02 \log M_\bullet/M_\odot + 36.35 ,
\end{equation}
where we note the near-linear scaling suggests an Eddington-limited energy source. 

A power-law relationship of this form allows a particularly simple calculation of the peak $g$-band luminosity function to be performed from our fitted black hole mass distribution, as 
\begin{equation}
{1 \over N_{\rm TDE}} {{\rm d} N_{\rm TDE} \over {\rm d} \log L_{g, {\rm peak}}} = {1 \over N_{\rm TDE}} {{\rm d} N_{\rm TDE} \over {\rm d} \log M_\bullet }  \times { {\rm d} \log M_\bullet \over {\rm d} \log L_{g, {\rm peak}}} ,
\end{equation}
where the final factor is a simple constant, and the black hole mass distribution must be evaluated at the black hole mass which corresponds to the specified peak $g$-band luminosity (using equation \ref{eq:lpeak}). As we have already derived the black hole mass distribution using the UV plateau data, this luminosity function is simple to compute. We can further compute the intrinsic peak optical $g$-band rate from 
\begin{equation}\label{eq:opt_rate}
{{\rm d} {\cal R} \over {\rm d} \log L_{g, {\rm peak}}} \propto  {1 \over L_{g, {\rm peak}}^{3/2}} {{\rm d} N_{\rm TDE} \over {\rm d} \log L_{g, {\rm peak}}}  ,
\end{equation}
where we use equation (\ref{eq:lpeak}) to evaluate the factor $L_{g, {\rm peak}}^{3/2}$.

The one non-trivial step in these calculations is to incorporate the effects of the Hills mass suppression at large black hole masses.  We do this using {\tt tidalspin}, the publicly available code produced in \cite{Mummery24}. This code convolves the naive $g$-band luminosity function derived above (which we denote $\bar p$) with the Hills suppression factor ${\cal H}$  
\begin{multline}
p(\log L_{g}) = \iiint \bar{p}(\log L_{g}) {\cal H}(a_\bullet, m_\star, \phi_{\rm orb}) \\ p(a_\bullet) p(m_\star) p(\phi_{\rm orb}) \, {\rm d}a_\bullet \, {\rm d}m_\star \, {\rm d}\phi_{\rm orb} ,
\end{multline}
where ${\cal H}=1$ if $M_\bullet < M_{\rm Hills}$ and ${\cal H}=0$ if $M_\bullet \geq M_{\rm Hills}$. The Hills mass is presented explicitly in Appendix \ref{app:stats}. The three priors $p(a_\bullet), p(m_\star), p(\phi_{\rm orb})$ are identical to those used in the UV plateau sampling procedure.  The effect of the Hills mass suppression is visible in the lowest panel of Figure \ref{fig:peakfuncs}, where we show by a red dotted line the peak $g$-band luminosity function in the absence of such an effect.  The only modifications to the rate occur at high $g$-band luminosities (which correspond to the high black hole masses close to the Hills mass).   The better agreement between the Hills-corrected luminosity function and the observational data can in some sense be taken as indirect evidence for the existence of event horizons in nature. 

Similarly to the X-ray analysis, the normalisation of the $g$-band luminosity rate is arbitrary (in our model), and was fit to match the amplitude of the \cite{Yao23} analysis. 

Just as with the peak X-ray luminosity function and the UV plateau luminosity function, this procedure is capable of reproducing the observed peak optical luminosity function. The null hypothesis that the observed and simulated distributions are the same cannot be rejected by a two sample KS test with $p$-value = 0.14.   This suggests that ultimately whatever is producing this early time emission, while potentially physically very complicated as a process, must on the output emission level be relatively simple, as it can be adequately described on the population level by a single near-linear relationship between $g$-band luminosity and black hole mass\footnote{If this relationship was taken to be strictly linear, then the tidal disruption event  $g$-band Eddington ratio  would be $L_{g, {\rm peak}}/ L_{\rm edd} \simeq 0.03$.}.

We note that, depending on how one calibrates the black hole masses of TDEs (i.e., with galactic scaling or the UV plateau relationship), different peak luminosity-black hole mass scaling relationships are found (L. Pouw et al., in prep.). We have found that varying the peak-mass relationship between these different scaling relationship calibrations does not qualitatively affect the results presented here. 
 
\section{Implications}\label{sec:implications}
\subsection{Tidal disruption event science}
We have demonstrated that it is possible to reproduce the tidal disruption event peak optical, late time UV and peak X-ray luminosity functions from first principles using models of relativistic accretion flows. The first, and most obvious, implication of these results is that the physics of relativistic accretion is essential to understanding the observed distributions of tidal disruption event properties. The second key implication of this work is the demonstration that the accretion flows which produce the late-time plateaus observed in optical/UV bands are the very same accretion flows which produce the early time X-ray luminosity observed in many of these systems. 

There are, however, a number of other implications of these results, which we take some time to discuss here. Firstly, this work demonstrates there is no ``missing energy'' in  tidal disruption event lightcurves. The so-called ``missing energy problem'' follows from the following observational statement, which was first discovered in the very early days of tidal disruption event surveys, that the integrated emitted energy in optical tidal disruption event light curves 
\begin{equation}
E_{\rm opt} = \int_0^\infty L_{\rm opt}(t) \, {\rm d}t, 
\end{equation}
is significant lower than the mass-energy available in the stellar debris $\sim M_\star c^2$. Typically, $E_{\rm opt} \sim 10^{50}$ erg, while the formal energy budget is $M_\star c^2 \sim 10^{54}$ erg, naively suggesting a very low radiative efficiency. However, every disc model used in this paper accretes at least $M_{\rm acc} \geq 0.04 M_\odot$ (half of the lowest mass star in our stellar mass function), with the usual accretion efficiency $\eta$ set exclusively by the black hole's spin. Ultimately therefore the solutions used in this work each release $\sim 10^{52}$ erg, as expected from simple energetic arguments. The vast majority of this radiated energy however is released in the form of (ultimately unobservable) extreme-UV photons, which  will be absorbed by neutral Hydrogen on their path to the observer. Further, much of this energy is released on decade timescales, not probed by the  analysis of the (non-disc) early optical emission.

In addition, while the physics at the heart of the models put forward here is that of the accretion process, which does not directly describe the luminosity observed at early times in optical bands, our results have a number of implications for the true physical origin of this emission component. Firstly, whatever its physical cause, the properties of this early time optical emission must be relatively simple on the population level. This is because the observed $g$-band luminosity can be adequately described, on the population level, by a single near-linear relationship between $g$-band luminosity and black hole mass. 

Furthermore, as the model put forward in this paper is able to reproduce the peak X-ray luminosity distribution (and these X-ray peaks typically occur at early times in the evolution of the system, when the early time optical component is still bright), our results seem to suggest that the early optical luminosity observed from these systems cannot be explained entirely by the reprocessing of X-ray emission to optical photons by a surrounding layer of material which persists for long (year) timescales. The reason for this is simple, if a significant fraction of the early time X-ray luminosity observed from these systems was being reprocessed, we would not be able to reproduce the observed X-ray luminosity function with a bare disc model (the disc model would systematically produce over-luminous X-ray emission). Reprocessing of X-ray emission could still be an acceptable model for the early time optical emission if it either (i) occurs over timescales shorter than the timescale at which the peak X-ray luminosity of the disc evolves, or (ii) acts in an all-or-nothing fashion, as a more gradual change in X-ray luminosity (either caused by partial reprocessing or reprocessing on longer timescales than the disc evolution) would alter the shape of the luminosity function (as opposed to simply reducing the amplitude at all luminosities, which appears to be the case). A third, possibly simpler, resolution of course is that the early time emission is not produced by the reprocessing of X-ray photons.   

Our results are suggestive, however, that some TDEs are not producing any observable X-ray emission for some reason which is not related to the physical properties (e.g., the temperature) of the accretion flow itself. This is related to the fact that at masses $M_\bullet \lesssim 10^6 M_\odot$ our model predicts nearly every TDE should be X-ray loud $L_X \gtrsim 10^{42}$ erg/s, in contention with the finding of \cite{Guolo24} although this was based on a small number of black hole mass measurements inferred from the $M_\bullet-\sigma$ relationship, and should be confirmed when larger data sets become available. Furthermore, 3/4 of our TDE sample (Fig. \ref{fig:xraysamp}) are predicted to be X-ray loud ($L_X\gtrsim 10^{42}$ erg/s), while only $\sim 40\%$ of observed optically-selected TDEs satisfy this criterion \cite{Guolo24}.  This may be a result of inefficient disc formation in some systems, which may also explain the lack of a detected UV plateau in $\sim 1/3$ of sources \cite{Mummery_et_al_2024}, or of course could simply be an observational selection effect \citep[the 40\% figure found by][is by definition a formal lower limit in this regard]{Guolo24}.  

Our final result relevant for tidal disruption event science is that our results suggest that the X-ray and optically selected tidal disruption event populations are drawn from the same mass distribution \citep[as was also argued by][based on the $M_\bullet-\sigma$ masses of the two populations]{Guolo24}. 

\subsection{Black hole physics}
\begin{figure}
     \centering
     \includegraphics[width=\linewidth]{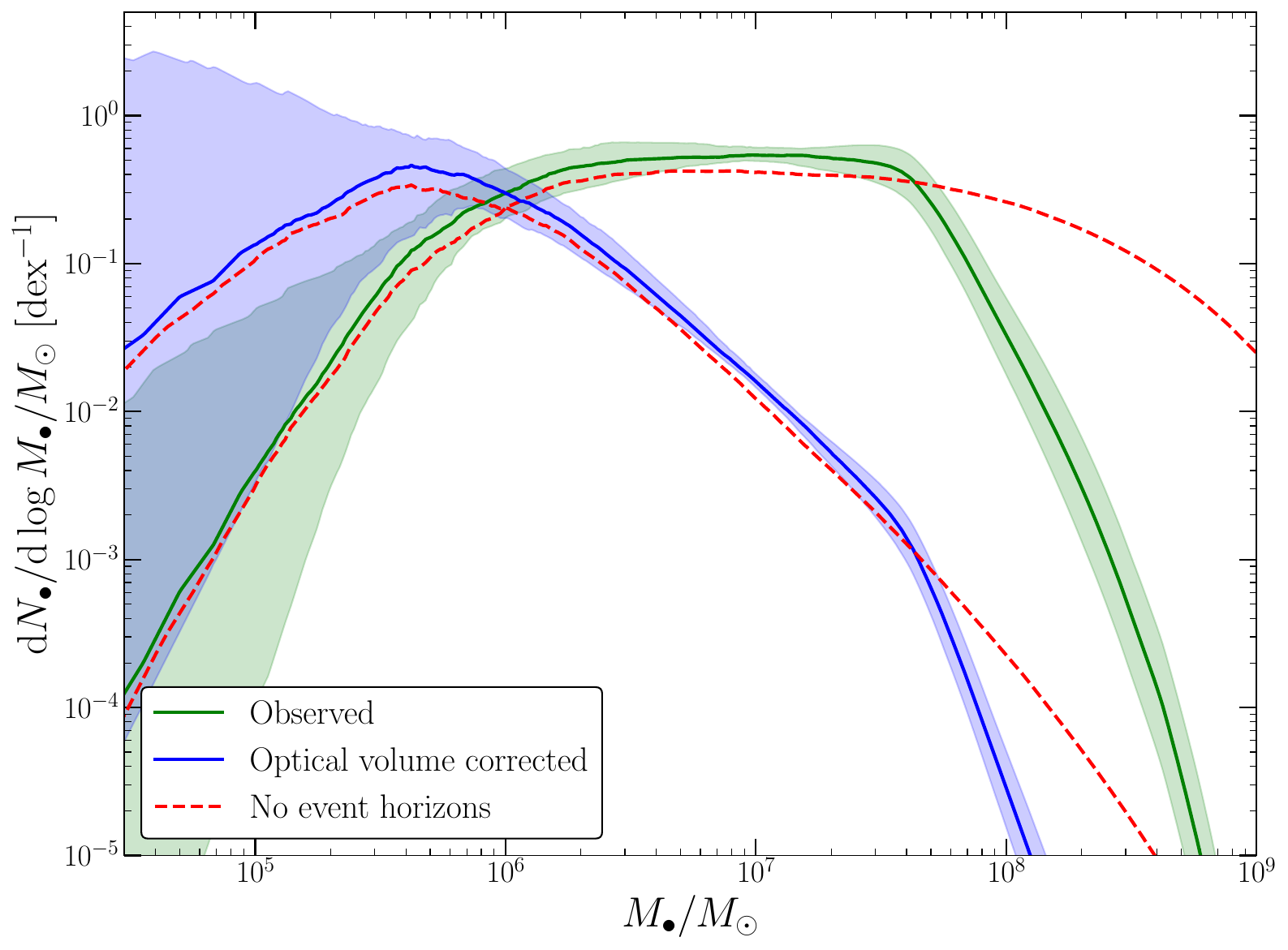}
     \caption{The black hole mass distributions implied by this work. In green we display the observed black hole mass distribution, while in blue we display this mass function once corrected for a simple Newtonian maximum observable volume effect (see text for further discussion). In both situations the solid curve shows the posterior median and the shaded region shows $1\sigma$ uncertainties. Red-dashed curves show the same posterior median curves when the effects of the Hills mass suppression are not taken into account. The normalisations of each curve are arbitrary, and were chosen so that the integral over all black hole masses of the green curves is unity.  We see that while there is a robust suppression in the observed TDE mass distribution below $10^6 M_\odot$, we cannot state with statistical significance that the volume-corrected mass distribution also shows a turn-over at these lower masses. Future  observations with surveys like Rubin/LSST will constrain the black hole mass function at these low masses.   }
     \label{fig:mass_dists}
 \end{figure}

In addition to the implications of these results for the physics of tidal disruption events themselves, we are now at a point in TDE science where we can start to make statements regarding the intrinsic black hole mass distribution in the local Universe. The parameterised black hole mass distribution we fit in this work is displayed in Figure \ref{fig:mass_dists}, along with a volume-corrected distribution. This volume correcting incorporates a  $1/{\cal V}_{\rm max} \propto L_{g, {\rm peak}}^{-3/2} \propto M_\bullet^{-1.53}$ factor to account for Malmquist bias (where we have used equation \ref{eq:lpeak} for the peak $g$-band luminosities dependence on black hole mass). In green we display the observed black hole mass distribution, while in blue we display this mass distribution once volume corrected. Red-dashed curves show the posterior median curves when the effects of the Hills mass suppression are not taken into account. We see that while there is a robust suppression in the observed TDE mass distribution below $10^6 M_\odot$, we cannot state with statistical significance that the volume-corrected mass distribution also shows a turn-over at these lower masses. Future  observations with surveys like Rubin/LSST will constrain the intrinsic black hole mass distribution at these low masses.

Note that this volume corrected distribution still does not necessarily correspond to the intrinsic black hole mass function,  as it does not take into account the intrinsic dependence of the TDE-rate on black hole mass. This intrinsic rate, which we denote ${\cal R}_{\rm TDE}(M_\bullet)$ quantifies the differing rate at which black holes of different  masses undergo tidal disruption events, with {\it all other parameters of the system fixed.} This factor can have a non-trivial dependence on black hole mass if, for example, lower mass black holes were to typically reside in more central concentrated galaxies which naturally produce a higher rate of tidal disruption events \citep[the consensus within the community is that indeed they do, e.g.,][]{Wang04}. The quantity we are able to constrain in this work is the product ${\rm d} N_\bullet / {\rm d}\log M_\bullet \propto {\cal R}_{\rm TDE}(M_\bullet)  \phi(M_\bullet)$, where $\phi(M_\bullet)$ is the intrinsic black hole mass function, a quantity of broad astrophysical interest. Most models for the mass dependence of ${\cal R}_{\rm TDE}$ are relatively weak, with \citep[e.g.][]{Alexander17, Yao23} suggesting  ${\cal R}_{\rm TDE} \propto M_\bullet^{-1/4}$, while \cite{Stone16} suggest a slightly steeper ${\cal R}_{\rm TDE} \propto M_\bullet^{-2/5}$, and \cite{Pfister20} suggest a slightly shallower ${\cal R}_{\rm TDE} \propto M_\bullet^{-0.14}$. In any of these cases the intrinsic black hole mass function $\phi(M_\bullet)$ would be qualitatively unchanged from that shown in Figure \ref{fig:mass_dists}, but more work \citep[both on the physics of ${\cal R}_{\rm TDE}$, e.g.,][and in extending the data sets used in this work]{Hannah24} is required to confirm this result.

\section{Conclusions}\label{sec:conclusions}
The key result of this paper is that the optical (Fig. \ref{fig:peakfuncs}), UV (Fig. \ref{fig:platfuncs}) and X-ray (Fig. \ref{fig:xrayfuncs}) properties of tidal disruption events can now be accurately described on the population level by using time-dependent relativistic models of accretion flows. The three free parameters fit in this work describe the observed tidal disruption event black hole mass distribution, and are of  astrophysical interest (e.g., Figure \ref{fig:mass_dists}). 

The mass and luminosity functions derived here will be of further use for predicting the future properties of (e.g.,) the optical/UV tidal disruption event population discovered by Rubin/LSST \citep[which is predicted to discover 10's of thousands of tidal disruption events, e.g.,][]{Bricman20}, or the X-ray properties of tidal disruption events discovered by the Einstein probe \citep[which is expected to discover a similar number to LSST][]{Yuan15}.  If future, better sampled, observational data sets provide even tighter constraints on the black hole mass function at the low mass end then this could have profound implications for the future LISA binary black hole detection rate \citep{Amaro-SeoaneLISA17}. 

The statistical framework developed in this paper can be extended more broadly to constrain other parameters of tidal disruption event systems. Of particular interest would be future black hole spin constrains which can be determined by the delicate interplay between observed tidal disruption event parameters and the Hills mass \citep[e.g.,][]{Mummery24}.

\section*{Acknowledgments} 
  
The authors would like to thank  E. Rossi and N. Stone for illuminating discussions during the early stages of this work.  This work was supported by a Leverhulme Trust International Professorship grant [number LIP-202-014]. For the purpose of Open Access, AM has applied a CC BY public copyright licence to any Author Accepted Manuscript version arising from this submission. This research was supported in part by grant no. NSF PHY-2309135 to the Kavli Institute for Theoretical Physics (KITP).

\section*{Data accessibility statement}
All optical/UV light curves and data products are available in the following repository  \href{https://github.com/sjoertvv/manyTDE}{https://github.com/sjoertvv/manyTDE}. The X-ray luminosity data was taken from  \cite{Guolo24}, and is publicly available. The peak optical luminosity function was taken from \cite{Yao23}, and is publicly available. The base {\tt FitTeD} code package used in this work is available to download at the following repository:  \href{https://bitbucket.org/fittingtransientswithdiscs/fitted_public/src}{https://bitbucket.org/fittingtransientswithdiscs/fitted\_public/src}.

\bibliographystyle{mnras}
\bibliography{andy}

\begin{thebibliography}{}
\makeatletter
\relax
\def\mn@urlcharsother{\let\do\@makeother \do\$\do\&\do\#\do\^\do\_\do\%\do\~}
\def\mn@doi{\begingroup\mn@urlcharsother \@ifnextchar [ {\mn@doi@} {\mn@doi@[]}}
\def\mn@doi@[#1]#2{\def\@tempa{#1}\ifx\@tempa\@empty \href {http://dx.doi.org/#2} {doi:#2}\else \href {http://dx.doi.org/#2} {#1}\fi \endgroup}
\def\mn@eprint#1#2{\mn@eprint@#1:#2::\@nil}
\def\mn@eprint@arXiv#1{\href {http://arxiv.org/abs/#1} {{\tt arXiv:#1}}}
\def\mn@eprint@dblp#1{\href {http://dblp.uni-trier.de/rec/bibtex/#1.xml} {dblp:#1}}
\def\mn@eprint@#1:#2:#3:#4\@nil{\def\@tempa {#1}\def\@tempb {#2}\def\@tempc {#3}\ifx \@tempc \@empty \let \@tempc \@tempb \let \@tempb \@tempa \fi \ifx \@tempb \@empty \def\@tempb {arXiv}\fi \@ifundefined {mn@eprint@\@tempb}{\@tempb:\@tempc}{\expandafter \expandafter \csname mn@eprint@\@tempb\endcsname \expandafter{\@tempc}}}

\bibitem[\protect\citeauthoryear{{Abramowicz}, {Czerny}, {Lasota}  \& {Szuszkiewicz}}{{Abramowicz} et~al.}{1988}]{Abram88}
{Abramowicz} M.~A.,  {Czerny} B.,  {Lasota} J.~P.,   {Szuszkiewicz} E.,  1988, \mn@doi [\apj] {10.1086/166683}, \href {https://ui.adsabs.harvard.edu/abs/1988ApJ...332..646A} {332, 646}

\bibitem[\protect\citeauthoryear{{Amaro-Seoane} et~al.,}{{Amaro-Seoane} et~al.}{2017}]{Amaro-SeoaneLISA17}
{Amaro-Seoane} P.,  et~al., 2017, \mn@doi [arXiv e-prints] {10.48550/arXiv.1702.00786}, \href {https://ui.adsabs.harvard.edu/abs/2017arXiv170200786A} {p. arXiv:1702.00786}

\bibitem[\protect\citeauthoryear{{Arcavi} et~al.,}{{Arcavi} et~al.}{2022a}]{Arcavi22TNS}
{Arcavi} I.,  et~al., 2022a, Transient Name Server Classification Report, \href {https://ui.adsabs.harvard.edu/abs/2022TNSCR.504....1A} {2022-504, 1}

\bibitem[\protect\citeauthoryear{{Arcavi}, {Dgany}, {Pellegrino}, {Howell}, {Burke}, {Gonzalez}, {Terreran}  \& {McCully}}{{Arcavi} et~al.}{2022b}]{Arcavi22TNSb}
{Arcavi} I.,  {Dgany} Y.,  {Pellegrino} C.,  {Howell} D.~A.,  {Burke} M.~N.~J.,  {Gonzalez} E.~P.,  {Terreran} G.,   {McCully} C.,  2022b, Transient Name Server Classification Report, \href {https://ui.adsabs.harvard.edu/abs/2022TNSCR.511....1A} {2022-511, 1}

\bibitem[\protect\citeauthoryear{{Balbus}}{{Balbus}}{2017}]{Balbus17}
{Balbus} S.~A.,  2017, \mn@doi [\mnras] {10.1093/mnras/stx1955}, \href {https://ui.adsabs.harvard.edu/abs/2017MNRAS.471.4832B} {471, 4832}

\bibitem[\protect\citeauthoryear{{Bricman} \& {Gomboc}}{{Bricman} \& {Gomboc}}{2020}]{Bricman20}
{Bricman} K.,  {Gomboc} A.,  2020, \mn@doi [\apj] {10.3847/1538-4357/ab6989}, \href {https://ui.adsabs.harvard.edu/abs/2020ApJ...890...73B} {890, 73}

\bibitem[\protect\citeauthoryear{{Charalampopoulos} et~al.,}{{Charalampopoulos} et~al.}{2024}]{Charalampopoulos24}
{Charalampopoulos} P.,  et~al., 2024, \mn@doi [arXiv e-prints] {10.48550/arXiv.2401.11773}, \href {https://ui.adsabs.harvard.edu/abs/2024arXiv240111773C} {p. arXiv:2401.11773}

\bibitem[\protect\citeauthoryear{{Dai}, {McKinney}, {Roth}, {Ramirez-Ruiz}  \& {Miller}}{{Dai} et~al.}{2018}]{Dai18}
{Dai} L.,  {McKinney} J.~C.,  {Roth} N.,  {Ramirez-Ruiz} E.,   {Miller} M.~C.,  2018, \mn@doi [\apjl] {10.3847/2041-8213/aab429}, \href {https://ui.adsabs.harvard.edu/abs/2018ApJ...859L..20D} {859, L20}

\bibitem[\protect\citeauthoryear{{Dexter} \& {Agol}}{{Dexter} \& {Agol}}{2009}]{DexterAgol09}
{Dexter} J.,  {Agol} E.,  2009, \mn@doi [\apj] {10.1088/0004-637X/696/2/1616}, \href {https://ui.adsabs.harvard.edu/abs/2009ApJ...696.1616D} {696, 1616}

\bibitem[\protect\citeauthoryear{{Done }, {Davis}, {Jin}, {Blaes}  \& {Ward}}{{Done } et~al.}{2012}]{Done12}
{Done } C.,  {Davis} S.~W.,  {Jin} C.,  {Blaes} O.,   {Ward} M.,  2012, \mn@doi [\mnras] {10.1111/j.1365-2966.2011.19779.x}, \href {https://ui.adsabs.harvard.edu/abs/2012MNRAS.420.1848D} {420, 1848}

\bibitem[\protect\citeauthoryear{{Eardley} \& {Lightman}}{{Eardley} \& {Lightman}}{1975}]{EardleyLightman75}
{Eardley} D.~M.,  {Lightman} A.~P.,  1975, \mn@doi [\apj] {10.1086/153777}, \href {https://ui.adsabs.harvard.edu/abs/1975ApJ...200..187E} {200, 187}

\bibitem[\protect\citeauthoryear{{Esquej} et~al.,}{{Esquej} et~al.}{2008}]{Esquej08}
{Esquej} P.,  et~al., 2008, \mn@doi [\aap] {10.1051/0004-6361:200810110}, \href {https://ui.adsabs.harvard.edu/abs/2008A&A...489..543E} {489, 543}

\bibitem[\protect\citeauthoryear{{Foreman-Mackey}, {Hogg}, {Lang}  \& {Goodman}}{{Foreman-Mackey} et~al.}{2013}]{EMCEE}
{Foreman-Mackey} D.,  {Hogg} D.~W.,  {Lang} D.,   {Goodman} J.,  2013, \mn@doi [\pasp] {10.1086/670067}, \href {https://ui.adsabs.harvard.edu/abs/2013PASP..125..306F} {125, 306}

\bibitem[\protect\citeauthoryear{{Fulton} et~al.,}{{Fulton} et~al.}{2022a}]{Fulton22TNS}
{Fulton} M.,  et~al., 2022a, Transient Name Server AstroNote, \href {https://ui.adsabs.harvard.edu/abs/2022TNSAN.106....1F} {106, 1}

\bibitem[\protect\citeauthoryear{{Fulton}, {Smith}, {Moore}, {Srivastav}  \& {Bruch}}{{Fulton} et~al.}{2022b}]{Fulton22TNSb}
{Fulton} M.,  {Smith} K.~W.,  {Moore} T.,  {Srivastav} S.,   {Bruch} R.~J.,  2022b, Transient Name Server Classification Report, \href {https://ui.adsabs.harvard.edu/abs/2022TNSCR.584....1F} {2022-584, 1}

\bibitem[\protect\citeauthoryear{{Greene}, {Strader}  \& {Ho}}{{Greene} et~al.}{2020}]{Greene20}
{Greene} J.~E.,  {Strader} J.,   {Ho} L.~C.,  2020, \mn@doi [\araa] {10.1146/annurev-astro-032620-021835}, \href {https://ui.adsabs.harvard.edu/abs/2020ARA&A..58..257G} {58, 257}

\bibitem[\protect\citeauthoryear{{Guillochon} \& {Ramirez-Ruiz}}{{Guillochon} \& {Ramirez-Ruiz}}{2013}]{Guillochon13}
{Guillochon} J.,  {Ramirez-Ruiz} E.,  2013, \mn@doi [\apj] {10.1088/0004-637X/767/1/25}, \href {https://ui.adsabs.harvard.edu/abs/2013ApJ...767...25G} {767, 25}

\bibitem[\protect\citeauthoryear{{Guolo}, {Gezari}, {Yao}, {van Velzen}, {Hammerstein}, {Cenko}  \& {Tokayer}}{{Guolo} et~al.}{2024}]{Guolo24}
{Guolo} M.,  {Gezari} S.,  {Yao} Y.,  {van Velzen} S.,  {Hammerstein} E.,  {Cenko} S.~B.,   {Tokayer} Y.~M.,  2024, \mn@doi [\apj] {10.3847/1538-4357/ad2f9f}, \href {https://ui.adsabs.harvard.edu/abs/2024ApJ...966..160G} {966, 160}

\bibitem[\protect\citeauthoryear{{Hammerstein}, {Gezari}, {Velzen}, {Kulkarni}, {Cenko}  \& {Yao}}{{Hammerstein} et~al.}{2021a}]{Hammerstein21TNS}
{Hammerstein} E.,  {Gezari} S.,  {Velzen} S.~V.,  {Kulkarni} S.,  {Cenko} B.,   {Yao} Y.,  2021a, Transient Name Server Classification Report, \href {https://ui.adsabs.harvard.edu/abs/2021TNSCR.732....1H} {2021-732, 1}

\bibitem[\protect\citeauthoryear{{Hammerstein}, {Gezari}, {Velzen}, {Kulkarni}, {Cenko}  \& {Yao}}{{Hammerstein} et~al.}{2021b}]{Hammerstein21TNSb}
{Hammerstein} E.,  {Gezari} S.,  {Velzen} S.~V.,  {Kulkarni} S.,  {Cenko} B.,   {Yao} Y.,  2021b, Transient Name Server Classification Report, \href {https://ui.adsabs.harvard.edu/abs/2021TNSCR.732....1H} {2021-732, 1}

\bibitem[\protect\citeauthoryear{{Hammerstein}, {Yao}, {Gezari}, {Velzen}, {Somalwar}  \& {Cenko}}{{Hammerstein} et~al.}{2022}]{Hammerstein22TNS}
{Hammerstein} E.,  {Yao} Y.,  {Gezari} S.,  {Velzen} S.~V.,  {Somalwar} J.,   {Cenko} B.,  2022, Transient Name Server Classification Report, \href {https://ui.adsabs.harvard.edu/abs/2022TNSCR.891....1H} {2022-891, 1}

\bibitem[\protect\citeauthoryear{{Holoien} et~al.,}{{Holoien} et~al.}{2016}]{Holoien16b}
{Holoien} T.~W.~S.,  et~al., 2016, \mn@doi [\mnras] {10.1093/mnras/stv2486}, \href {https://ui.adsabs.harvard.edu/abs/2016MNRAS.455.2918H} {455, 2918}

\bibitem[\protect\citeauthoryear{{Ingram}, {Mastroserio}, {Dauser}, {Hovenkamp}, {van der Klis}  \& {Garc{\'\i}a}}{{Ingram} et~al.}{2019}]{Ingram19}
{Ingram} A.,  {Mastroserio} G.,  {Dauser} T.,  {Hovenkamp} P.,  {van der Klis} M.,   {Garc{\'\i}a} J.~A.,  2019, \mn@doi [\mnras] {10.1093/mnras/stz1720}, \href {https://ui.adsabs.harvard.edu/abs/2019MNRAS.488..324I} {488, 324}

\bibitem[\protect\citeauthoryear{{Kippenhahn} \& {Weigert}}{{Kippenhahn} \& {Weigert}}{1990}]{Kippenhahn90}
{Kippenhahn} R.,  {Weigert} A.,  1990, {Stellar Structure and Evolution}

\bibitem[\protect\citeauthoryear{{Kroupa}}{{Kroupa}}{2001}]{Kroupa01}
{Kroupa} P.,  2001, \mn@doi [\mnras] {10.1046/j.1365-8711.2001.04022.x}, \href {https://ui.adsabs.harvard.edu/abs/2001MNRAS.322..231K} {322, 231}

\bibitem[\protect\citeauthoryear{{Lynden-Bell} \& {Pringle}}{{Lynden-Bell} \& {Pringle}}{1974}]{LBP74}
{Lynden-Bell} D.,  {Pringle} J.~E.,  1974, \mn@doi [MNRAS] {10.1093/mnras/168.3.603}, \href {https://ui.adsabs.harvard.edu/abs/1974MNRAS.168..603L} {168, 603}

\bibitem[\protect\citeauthoryear{{Magorrian} \& {Tremaine}}{{Magorrian} \& {Tremaine}}{1999}]{Magorrian99}
{Magorrian} J.,  {Tremaine} S.,  1999, \mn@doi [\mnras] {10.1046/j.1365-8711.1999.02853.x}, \href {https://ui.adsabs.harvard.edu/abs/1999MNRAS.309..447M} {309, 447}

\bibitem[\protect\citeauthoryear{{Metzger}}{{Metzger}}{2022}]{Metzger22}
{Metzger} B.~D.,  2022, \mn@doi [\apjl] {10.3847/2041-8213/ac90ba}, \href {https://ui.adsabs.harvard.edu/abs/2022ApJ...937L..12M} {937, L12}

\bibitem[\protect\citeauthoryear{{Mummery}}{{Mummery}}{2021}]{Mum21}
{Mummery} A.,  2021, \mn@doi [\mnras] {10.1093/mnras/stab1187}, \href {https://ui.adsabs.harvard.edu/abs/2021MNRAS.504.5144M} {504, 5144}

\bibitem[\protect\citeauthoryear{{Mummery}}{{Mummery}}{2023}]{Mummery23a}
{Mummery} A.,  2023, \mn@doi [\mnras] {10.1093/mnras/stac2846}, \href {https://ui.adsabs.harvard.edu/abs/2023MNRAS.518.1905M} {518, 1905}

\bibitem[\protect\citeauthoryear{{Mummery}}{{Mummery}}{2024}]{Mummery24}
{Mummery} A.,  2024, \mn@doi [\mnras] {10.1093/mnras/stad3636}, \href {https://ui.adsabs.harvard.edu/abs/2024MNRAS.527.6233M} {527, 6233}

\bibitem[\protect\citeauthoryear{{Mummery} \& {Balbus}}{{Mummery} \& {Balbus}}{2020}]{MumBalb20a}
{Mummery} A.,  {Balbus} S.~A.,  2020, \mn@doi [\mnras] {10.1093/mnras/staa192}, \href {https://ui.adsabs.harvard.edu/abs/2020MNRAS.492.5655M} {492, 5655}

\bibitem[\protect\citeauthoryear{{Mummery}, {Wevers}, {Saxton}  \& {Pasham}}{{Mummery} et~al.}{2023}]{Mummery_Wevers_23}
{Mummery} A.,  {Wevers} T.,  {Saxton} R.,   {Pasham} D.,  2023, \mn@doi [\mnras] {10.1093/mnras/stac3798}, \href {https://ui.adsabs.harvard.edu/abs/2023MNRAS.519.5828M} {519, 5828}

\bibitem[\protect\citeauthoryear{{Mummery}, {Nathan}, {Ingram}  \& {Gardner}}{{Mummery} et~al.}{2024a}]{mummery2024fitted}
{Mummery} A.,  {Nathan} E.,  {Ingram} A.,   {Gardner} M.,  2024a, arXiv e-prints, \href {https://ui.adsabs.harvard.edu/abs/2024arXiv240815048M} {p. arXiv:2408.15048}

\bibitem[\protect\citeauthoryear{{Mummery}, {van Velzen}, {Nathan}, {Ingram}, {Hammerstein}, {Fraser-Taliente}  \& {Balbus}}{{Mummery} et~al.}{2024b}]{Mummery_et_al_2024}
{Mummery} A.,  {van Velzen} S.,  {Nathan} E.,  {Ingram} A.,  {Hammerstein} E.,  {Fraser-Taliente} L.,   {Balbus} S.,  2024b, \mn@doi [\mnras] {10.1093/mnras/stad3001}, \href {https://ui.adsabs.harvard.edu/abs/2024MNRAS.527.2452M} {527, 2452}

\bibitem[\protect\citeauthoryear{{Nicholl} et~al.,}{{Nicholl} et~al.}{2019}]{Nicholl19}
{Nicholl} M.,  et~al., 2019, \mn@doi [\mnras] {10.1093/mnras/stz1837}, \href {https://ui.adsabs.harvard.edu/abs/2019MNRAS.488.1878N} {488, 1878}

\bibitem[\protect\citeauthoryear{{Piran}, {Svirski}, {Krolik}, {Cheng}  \& {Shiokawa}}{{Piran} et~al.}{2015}]{Piran15}
{Piran} T.,  {Svirski} G.,  {Krolik} J.,  {Cheng} R.~M.,   {Shiokawa} H.,  2015, \mn@doi [\apj] {10.1088/0004-637X/806/2/164}, \href {https://ui.adsabs.harvard.edu/abs/2015ApJ...806..164P} {806, 164}

\bibitem[\protect\citeauthoryear{{Rees}}{{Rees}}{1988}]{Rees88}
{Rees} M.~J.,  1988, \mn@doi [\nat] {10.1038/333523a0}, \href {https://ui.adsabs.harvard.edu/abs/1988Natur.333..523R} {333, 523}

\bibitem[\protect\citeauthoryear{{Ryu}, {Krolik}  \& {Piran}}{{Ryu} et~al.}{2020}]{Ryu20}
{Ryu} T.,  {Krolik} J.,   {Piran} T.,  2020, \mn@doi [\apj] {10.3847/1538-4357/abbf4d}, \href {https://ui.adsabs.harvard.edu/abs/2020ApJ...904...73R} {904, 73}

\bibitem[\protect\citeauthoryear{{Sazonov} et~al.,}{{Sazonov} et~al.}{2021}]{Sazonov21}
{Sazonov} S.,  et~al., 2021, \mn@doi [\mnras] {10.1093/mnras/stab2843}, \href {https://ui.adsabs.harvard.edu/abs/2021MNRAS.508.3820S} {508, 3820}

\bibitem[\protect\citeauthoryear{{Schechter}}{{Schechter}}{1976}]{Schechter76}
{Schechter} P.,  1976, \mn@doi [\apj] {10.1086/154079}, \href {https://ui.adsabs.harvard.edu/abs/1976ApJ...203..297S} {203, 297}

\bibitem[\protect\citeauthoryear{{Shakura} \& {Sunyaev}}{{Shakura} \& {Sunyaev}}{1973}]{SS73}
{Shakura} N.~I.,  {Sunyaev} R.~A.,  1973, \aap, \href {https://ui.adsabs.harvard.edu/abs/1973A&A....24..337S} {24, 337}

\bibitem[\protect\citeauthoryear{{Shankar}, {Salucci}, {Granato}, {De Zotti}  \& {Danese}}{{Shankar} et~al.}{2004}]{Shankar04}
{Shankar} F.,  {Salucci} P.,  {Granato} G.~L.,  {De Zotti} G.,   {Danese} L.,  2004, \mn@doi [\mnras] {10.1111/j.1365-2966.2004.08261.x}, \href {https://ui.adsabs.harvard.edu/abs/2004MNRAS.354.1020S} {354, 1020}

\bibitem[\protect\citeauthoryear{{Somalwar} et~al.,}{{Somalwar} et~al.}{2023}]{Somalwar23}
{Somalwar} J.~J.,  et~al., 2023, \mn@doi [arXiv e-prints] {10.48550/arXiv.2310.03791}, \href {https://ui.adsabs.harvard.edu/abs/2023arXiv231003791S} {p. arXiv:2310.03791}

\bibitem[\protect\citeauthoryear{{Stone} \& {Metzger}}{{Stone} \& {Metzger}}{2016}]{Stone16}
{Stone} N.~C.,  {Metzger} B.~D.,  2016, \mn@doi [\mnras] {10.1093/mnras/stv2281}, \href {https://ui.adsabs.harvard.edu/abs/2016MNRAS.455..859S} {455, 859}

\bibitem[\protect\citeauthoryear{{Wang} \& {Merritt}}{{Wang} \& {Merritt}}{2004}]{Wang04}
{Wang} J.,  {Merritt} D.,  2004, \mn@doi [\apj] {10.1086/379767}, \href {https://ui.adsabs.harvard.edu/abs/2004ApJ...600..149W} {600, 149}

\bibitem[\protect\citeauthoryear{{Wen}, {Jonker}, {Stone}, {Zabludoff}  \& {Psaltis}}{{Wen} et~al.}{2020}]{Wen20}
{Wen} S.,  {Jonker} P.~G.,  {Stone} N.~C.,  {Zabludoff} A.~I.,   {Psaltis} D.,  2020, \mn@doi [\apj] {10.3847/1538-4357/ab9817}, \href {https://ui.adsabs.harvard.edu/abs/2020ApJ...897...80W} {897, 80}

\bibitem[\protect\citeauthoryear{{Wen}, {Jonker}, {Stone}  \& {Zabludoff}}{{Wen} et~al.}{2021}]{Wen21}
{Wen} S.,  {Jonker} P.~G.,  {Stone} N.~C.,   {Zabludoff} A.~I.,  2021, \mn@doi [\apj] {10.3847/1538-4357/ac00b5}, \href {https://ui.adsabs.harvard.edu/abs/2021ApJ...918...46W} {918, 46}

\bibitem[\protect\citeauthoryear{{Wevers} et~al.,}{{Wevers} et~al.}{2021}]{Wevers21}
{Wevers} T.,  et~al., 2021, \mn@doi [\apj] {10.3847/1538-4357/abf5e2}, \href {https://ui.adsabs.harvard.edu/abs/2021ApJ...912..151W} {912, 151}

\bibitem[\protect\citeauthoryear{{Wevers}, {Guolo}, {Pasham}, {Coughlin}, {Tombesi}, {Yao}  \& {Gezari}}{{Wevers} et~al.}{2024}]{Wevers24}
{Wevers} T.,  {Guolo} M.,  {Pasham} D.~R.,  {Coughlin} E.~R.,  {Tombesi} F.,  {Yao} Y.,   {Gezari} S.,  2024, \mn@doi [\apj] {10.3847/1538-4357/ad1878}, \href {https://ui.adsabs.harvard.edu/abs/2024ApJ...963...75W} {963, 75}

\bibitem[\protect\citeauthoryear{{Yang} \& {Wang}}{{Yang} \& {Wang}}{2013}]{YangWang13}
{Yang} X.,  {Wang} J.,  2013, \mn@doi [\apjs] {10.1088/0067-0049/207/1/6}, \href {https://ui.adsabs.harvard.edu/abs/2013ApJS..207....6Y} {207, 6}

\bibitem[\protect\citeauthoryear{{Yao}, {Graham}, {Rodriguez}, {Somalwar}, {Velzen}, {Gezari}  \& {Hammerstein}}{{Yao} et~al.}{2022}]{Yao22TNS}
{Yao} Y.,  {Graham} M.,  {Rodriguez} A.,  {Somalwar} J.,  {Velzen} S.~V.,  {Gezari} S.,   {Hammerstein} E.,  2022, Transient Name Server Classification Report, \href {https://ui.adsabs.harvard.edu/abs/2022TNSCR.620....1Y} {2022-620, 1}

\bibitem[\protect\citeauthoryear{{Yao} et~al.,}{{Yao} et~al.}{2023}]{Yao23}
{Yao} Y.,  et~al., 2023, \mn@doi [\apjl] {10.3847/2041-8213/acf216}, \href {https://ui.adsabs.harvard.edu/abs/2023ApJ...955L...6Y} {955, L6}

\bibitem[\protect\citeauthoryear{{Yuan} et~al.,}{{Yuan} et~al.}{2015}]{Yuan15}
{Yuan} W.,  et~al., 2015, arXiv e-prints, \href {https://ui.adsabs.harvard.edu/abs/2015arXiv150607735Y} {p. arXiv:1506.07735}

\bibitem[\protect\citeauthoryear{{van Velzen} et~al.,}{{van Velzen} et~al.}{2019a}]{vanVelzen19b}
{van Velzen} S.,  et~al., 2019a, \mn@doi [\apj] {10.3847/1538-4357/aafe0c}, \href {https://ui.adsabs.harvard.edu/abs/2019ApJ...872..198V} {872, 198}

\bibitem[\protect\citeauthoryear{{van Velzen}, {Stone}, {Metzger}, {Gezari}, {Brown}  \& {Fruchter}}{{van Velzen} et~al.}{2019b}]{vanVelzen19}
{van Velzen} S.,  {Stone} N.~C.,  {Metzger} B.~D.,  {Gezari} S.,  {Brown} T.~M.,   {Fruchter} A.~S.,  2019b, \mn@doi [\apj] {10.3847/1538-4357/ab1844}, \href {https://ui.adsabs.harvard.edu/abs/2019ApJ...878...82V} {878, 82}

\bibitem[\protect\citeauthoryear{{van Velzen} et~al.,}{{van Velzen} et~al.}{2021}]{vanVelzen21}
{van Velzen} S.,  et~al., 2021, \mn@doi [\apj] {10.3847/1538-4357/abc258}, \href {https://ui.adsabs.harvard.edu/abs/2021ApJ...908....4V} {908, 4}

\makeatother
\end{thebibliography}

\appendix{}
\onecolumn
\section{Statistical Formalism}\label{app:stats}
In this Appendix we shall introduce and formalise the statistical procedure used in this paper to derive constraints on the black hole mass distribution of those black holes represented in the tidal disruption event population. In deriving this formalism it will prove most useful to consider a more general approach than is strictly required for the present calculations, as the procedure used in this paper can be generalised to study the distributions of other key parameters (i.e., the stellar properties of the tidal disruption event population), this may be of interest in the future when more data is available.

To keep this formalism general, let us suppose there is a function $f$ which returns a tidal disruption event observable $L$ from a set of physical input parameters $\left\{ \Psi \right\}$, i.e., 
\begin{equation}
    L = f(\left\{ \Psi \right\}) .
\end{equation}
The notation $L$ is used here as the observable is most likely to be some observed luminosity (but in principle this could be any quantity), and the list of input parameters is formally completely general but is likely to be (at least for the accretion problem at hand)
\begin{equation}
    \left\{ \Psi \right\} = \left[ M_\bullet, a_\bullet, M_\star, R_\star , \theta_{\rm obs}, \phi_{\rm orb}, \alpha, \beta, \dots \right]. 
\end{equation}
Here notation is completely standard: $M_\bullet, a_\bullet$ are the black hole mass and spin; $M_\star, R_\star$ the stellar mass and radius; $\theta_{\rm obs}$ is the disc-observer inclination and $\phi_{\rm orb}$ is the angle made between the incoming stellar orbit and the black hole's spin axis (which is an important parameter for determining the Hills mass); $\alpha$ represents the classical disc turbulence parameter of \cite{SS73}; while $\beta$ is the impact parameter. Naturally, this list could be extended if required (e.g., to include stellar metalicity, etc.).  

Further suppose that the probability distributions of each of our parameters is known (or at least a suitable parameterised distribution can be written down). We will use the following notation for all probability density functions: $p_X(x){\rm d}x$ is the probability that variable $X$ takes a value in the range $x \to x+{\rm d}x$.  It follows that the probability density function for our observable $L$ is given by 
\begin{equation}
    p_L(l) \propto \iint \dots \int p_{\Psi} (\psi) \,  \delta\left(l - f(\psi)\right) \, {\rm d}^N \psi , 
\end{equation}
where the integral is over all $N$ variables in the list $\left\{ \Psi \right\}$, and the function $p_\Psi(\psi)$ is the {\it joint} probability density function of the $N$ variables. The normalisation of this proportionality, which is not required for the calculation at hand, is given by the same integral as above but with an additional integration over all observables $l$.  In words this simply states that the probability of observing a given value $l$ of the observable $L$ is the sum of all possible ways of obtaining that value $l$ with different input parameters, weighted by the relative probabilities of those input parameters occurring.   To be explicit, for a tidal disruption event this would look something like (where `$\dots$' indicate other possible parameters to be marginalised over) 
\begin{multline}
        p_L(l) \propto \int p_{M_\bullet}(m_\bullet) \int p_{A_\bullet}(a_\bullet) \int p_{M_\star }(m_\star ) \int p_{\Theta_{\rm obs}}(\theta_{\rm obs})  \int p_{\Phi_{\rm orb}}(\phi_{\rm orb}) \dots \,\,  \\ \Theta\left(M_{\rm hills}(a_\bullet, m_\star, \phi_{\rm orb},  \dots)  - m_\bullet\right) \delta\left(l - f(m_\bullet, a_\bullet, m_\star, \theta_{\rm obs}, \dots )\right) \, {\rm d}m_\bullet \, {\rm d}a_\bullet \, {\rm d}m_\star \, {\rm d}\theta_{\rm obs} \, {\rm d}\phi_{\rm orb} \dots 
\end{multline}
The above integral introduces the first non-trivial point of statistics regarding the tidal disruption event calculation. Even if the parameter distributions of (e.g.)  black hole masses and spins are independent in the wider universe they become fundamentally coupled when considering tidal disruption events as only certain sets of parameters -- those which produce a tidal radius outside of the event horizon -- are observable (the fact that the statistical independence of black hole masses and spins is unlikely to be true in nature is beside the point). This coupling can be simply modelled by the inclusion of a Heaviside theta function, defined as
\begin{equation}
    \Theta\left(M_{\rm hills}(a_\bullet, m_\star, \phi_{\rm orb}, \dots)  - m_\bullet\right) = 
    \begin{cases}
        1, \quad {\rm if} \,\, M_{\rm hills}(a_\bullet, m_\star, \phi_{\rm orb}, \dots)  > m_\bullet, \\
        \\
        0, \quad {\rm if} \,\, M_{\rm hills}(a_\bullet, m_\star, \phi_{\rm orb}, \dots)  \leq m_\bullet .
    \end{cases}
\end{equation}
The so-called Hills mass $M_{\rm Hills}$ represents the limiting black hole mass at which (for a given stellar mass and radius) the tidal disruption occurs precisely at the point at which all of the debris is immediately lost within the event horizon (and therefore no observable emission can be detected). The Hills mass can be derived analytically in the Kerr metric, as was recently shown by \cite{Mummery24}, and is given by 
\begin{equation}\label{HillsMass}
      M_{\rm Hills}(a_\bullet, \phi_{\rm orb}, M_\star, R_\star) = \left[{2 R_\star^3 c^6 \over  G^3 M_\star } \right]^{1/2} {1\over \chi^{3/2}} 
     \Bigg[1 + {6\chi \over \chi^2 - a_\bullet^2 \cos^2\phi_{\rm orb}} + {3a_\bullet^2 \over 2\chi^2} - {6a_\bullet \sin \phi_{\rm orb} \over \sqrt{\chi^3 - a_\bullet^2 \chi \cos^2\phi_{\rm orb}}} \Bigg]^{1/2}, 
\end{equation}
where  $\chi(a_\bullet, \phi_{\rm orb})$ is the root of 
\begin{equation}
    \chi^4 - 4\chi^3 - a_\bullet^2(1 - 3 \cos^2\phi_{\rm orb})\chi^2 + a_\bullet^4\cos^2\phi_{\rm orb} + 4a_\bullet \sin \phi_{\rm orb} \sqrt{\chi^5 - a_\bullet^2\chi^3\cos^2\phi_{\rm orb}} = 0 .
\end{equation}
Including a Heaviside theta function in this manner allows each parameter integral to be performed independently. This is done for convenience for later steps and does not affect the statistical formalism.  

Now, as written the probability density function of each variable only depends on the value that variable takes. This would be true if we had a perfect description of the underlying distributions, with no tuneable parameters.  Of course, in reality, we are in fact interested only in the values that these tuneable parameters take. To bring this formalism back in line with the main body of the paper we shall use the black hole mass distribution as an explicit example. The simple functional form of the black hole mass distribution used in this work is 
\begin{equation}
    p_{M_\bullet}(m_\bullet; \alpha_h, \alpha_l, M_c, M_g, \gamma) \propto  {m_\bullet^{\alpha_h} \over  1 + (M_c/m_\bullet)^{\alpha_l-\alpha_h} } \exp\left(- \left({m_\bullet\over M_g}\right)^\gamma\right) .
\end{equation}
The point being that in reality each of our probability density functions has tuneable parameters ($\alpha_h, \alpha_l, M_c, M_g, \gamma$ in the above), and our probability density function for our observable $L$ is a function of {\it only} these parameters. Call the list of all tuneable parameters $\left\{ \xi_i \right\}$, then 
\begin{equation}
    p_L(l ,\left\{ \xi_i \right\}) \propto \iint \dots \int p_{\Psi} (\psi, \left\{ \xi_i \right\}) \,  \delta\left(l - f(\psi)\right) \, {\rm d}^N \psi .
\end{equation}
This construction allows us to determine $\left\{ \xi_i \right\}$ given a set of observations $\left\{ L_j \right\}$. By definition, the likelihood of observing $\left\{ L_j \right\}$ is 
\begin{equation}
    {\cal L}\left(\left\{ \xi_i \right\}\right) = \prod_{j =1}^{N_{\rm TDE}} \, p_L\left(L_j, \left\{ \xi_i \right\} \right) , 
\end{equation}
which is a function only of $\left\{ \xi_i\right\}$ (i.e., the likelihood is formally the probability of observing the set $\left\{ L_j \right\}$ given a probability density function $p_L(l, \left\{ \xi_i \right\})$). Of course one should work with the log likelihood as this is much less likely to be affected by numerical rounding errors 
\begin{equation}
    \log {\cal L}\left(\left\{ \xi_i \right\}\right) = \sum_{j =1}^{N_{\rm TDE}} \, \log p_L\left(L_j, \left\{ \xi_i \right\} \right) . 
\end{equation}
Then, given a set of observations $\left\{ L_j \right\}$, the maximum likelihood estimation of the underlying tuneable parameters $\left\{ \xi_i \right\}$ is just 
\begin{equation}
    \left\{ \xi_\star \right\} = \max_{\left\{ \xi_i \right\}} \, \Big[\log {\cal L}\left(\left\{ \xi_i \right\}\right) \Big]. 
\end{equation}
The parameter set $\left\{ \xi_\star \right\}$ is of obvious astrophysical interest, and is what we solve for in this work.  

\subsection{Weighted likelihood}
The likelihood constructed above will, when suitably maximised, return the tuneable parameters $\left\{\xi_i\right\}$ which best describe the observed distribution of tidal disruption parameters. While this is the quantity which we work with in this paper it is not always the distribution of most interest. For other applications, we are often more interested in the parameter distribution which best describes the true intrinsic {\it rate} at which different tidal disruption event properties occur in the wider Universe, and not the sub-set of which we are ultimately able to observe. To complete the statistical formalism presented in this Appendix, we describe how that approach would be performed. 

To constrain the intrinsic distribution the likelihood derived above must be re-weighted, in the following manner.  With each tidal disruption event in our sample we can associate a volume ${\cal V}_j$, which is given by the maximum volume out to which this particular tidal disruption event could have been observed. Naturally, sources with larger/smaller ${\cal V}_j$ will be over/under represented in our sample, compared to their true intrinsic rates. The likelihood therefore must, if the goal is to understand the intrinsic rates of properties of tidal disruption events, be weighted by the inverse source volume of each tidal disruption event. Defining $\widehat {\cal L}$ as this weighted likelihood, the weighting must be performed in the following manner 
\begin{equation}
    \widehat {\cal L}\left(\left\{ \xi_i \right\}\right) = \prod_{j =1}^{N_{\rm TDE}} \, \left[p_L\left(L_j, \left\{ \xi_i \right\} \right)\right]^{w_j} , 
\end{equation}
where $w_j \propto 1/{\cal V}_j$. In words, the mathematical reason for this is that a source in our sample with a volume $1/8^{\rm th}$ of another will in reality occur 8 times as frequently in the global population (and should therefore be counted 8 times). In log-space we therefore have 
\begin{equation}
    \log \widehat {\cal L}\left(\left\{ \xi_i \right\}\right) = \sum_{j =1}^{N_{\rm TDE}} \, w_j   \log p_L\left(L_j, \left\{ \xi_i \right\} \right) . 
\end{equation}
The volume ${\cal V}_j$ can be simply estimated for each source in our sample as our tidal disruption event population is constructed from optical survey telescopes, with a known flux detection limit, $F_{\rm lim}$. In a Newtonian cosmology (which is a reasonable approximation for nearly all tidal disruption events which are found at $z \lesssim 0.1$, although we note that our most luminous TDEs are observed out to $z\sim 0.3$), the maximum distance out to which a source with peak optical luminosity $L_{g, {\rm peak}, j}$ can be observed is 
\begin{equation}
    D_{\rm max} = \sqrt{L_{g, {\rm peak}, j} \over 4 \pi F_{\rm lim}}. 
\end{equation}
Here $L_{\rm peak}$ is measured at $(1+z)$ times the frequency that corresponds to the $F_{\rm lim}$. With $z<0.1$, this so-called $K$-correction is a weak function of redshift and we therefore obtain a simple estimate of the volume out to which each tidal disruption event can be detected:
\begin{equation}
    {\cal V}_j \propto D^3_{\rm max} \propto \left(L_{g, {\rm peak}, j}\right)^{3/2} . 
\end{equation}
As the absolute value of the likelihood is not of interest to this calculation, one can define likelihood weights for each tidal disruption event with respect to (e.g.) a reference luminosity $L_0 = 10^{43}$ erg/s, or explicitly 
\begin{equation}
    w_j \propto {1\over {\cal V}_j} = \left({L_{g, {\rm peak}, j} \over L_0}\right)^{-3/2} .
\end{equation}

\subsection{Practical implementation of formalism }\label{implimentation}
Given a set of starting distributions with tuneable parameters $\left\{ \xi_i \right\}$, and an observational set $\left\{ L_j \right\}$, the only non-trivial step actually required in computing the likelihood is performing the integral 
\begin{equation}
    p_L(l ,\left\{ \xi_i \right\}) \propto \iint \dots \int p_{\Psi} (\psi, \left\{ \xi_i \right\}) \,  \delta\left(l - f(\psi)\right) \, {\rm d}^N \psi .
\end{equation}
For the problem at hand we are interested in constraining parameters only in the black hole mass distribution. We are interested therefore in solving the integral 
\begin{multline}
        p_L(l, \left\{ \xi_i \right\}) \propto \int p_{M_\bullet}(m_\bullet , \left\{ \xi_{i} \right\}) \int p_{A_\bullet}(a_\bullet) \int p_{M_\star }(m_\star) \int p_{\Theta_{\rm obs}}(\theta_{\rm obs}) \int p_{\Phi_{\rm orb}}(\phi_{\rm orb}) \dots  \\ \,\,  \Theta\left(M_{\rm hills}(a_\bullet, m_\star, \phi_{\rm orb}, \dots)  - m_\bullet\right) \delta\left(l - f(m_\bullet, a_\bullet, m_\star, \theta_{\rm obs}, \dots )\right) \, {\rm d}m_\bullet \, {\rm d}a_\bullet \, {\rm d}m_\star \, {\rm d}\theta_{\rm obs} \, {\rm d}\phi_{\rm orb} \dots ,
\end{multline}
Computing this integral is non-trivial for two main reasons: the coupling between parameters introduced by the Hills mass and the non-analytical nature of the plateau luminosity function $f$ (the plateau luminosity for a given parameter set $\left\{ \Psi \right\}$ is computed numerically). Even if we were to analytically write down an approximation for $f$, the Hills mass coupling still makes this integral non-trivial to compute analytically. We therefore turn to a numerical approach to approximating this integral. 

Fortunately, both the Heaviside theta function (modelling the Hills mass) and the delta function act only as simple binary switches,  and this integral is therefore easily approximated by Monte Carlo techniques.   Formally, in the limit $N_{\rm sample} \to \infty$, the following procedure (written in pseudo-code) asymptotes exactly to $p_L$: 
\begin{itemize}
    \item Sample a set of parameters from each of the specified distributions $m_{\bullet, k} \sim p_{M_\bullet}(m_\bullet, \left\{ \xi_{i} \right\})$, $m_{\star, k} \sim p_{M_\star}(m_\star), \dots$ \\ 
    \item Check if $M_{\rm hills}(a_{\bullet, k}, m_{\star, k}, \dots)  > m_{\bullet, k}$ \\
    \item If so, compute $l_k = f\left(m_{\bullet, k}, a_{\bullet, k}, m_{\star, k}, \theta_k, \dots \right)$ \\
    \item Repeat $k = 1, 2, 3, \dots, N_{\rm sample}$ times \\
    \item Construct a normalised histogram $\widetilde p_L(l, \left\{ \xi_i \right\}; N_{\rm sample})$ of the $\left\{ l_k \right\}$ luminosities  \\
    \item In the limit $N_{\rm sample} \to \infty$, we have the following exact limit $\widetilde p_L(l, \left\{ \xi_i \right\}; N_{\rm sample})\to p_L(l, \left\{ \xi_i \right\})$ \\
    \item For large finite $N_{\rm sample}$, the function $\widetilde p_L(l, \left\{ \xi_i \right\}; N_{\rm sample})$ will be a good approximation of the solutions to the required integrals \\
    \item One can then approximate $\log \widetilde {\cal L}\left(\left\{ \xi_i \right\}\right) = \sum_{j =1}^{N_{\rm TDE}} \, \log \widetilde p_L\left(L_j, \left\{ \xi_i \right\}; N_{\rm sample} \right)$, and perform standard techniques to fit for $ \left\{ \xi_\star \right\}$
\end{itemize}
By experimenting numerically we found that $N_{\rm sample} = 10^4$ was sufficient to remove noticeable stochasticity in $\widetilde p_L$. To avoid zero probability cases resulting from the finite $N_{\rm sample}$ used in this work, we smooth the luminosity histogram $\widetilde p_L$ by using Gaussian kernel density estimation. 

\twocolumn
\section{TDE discovery references}\label{app:discovery}

\renewcommand{\arraystretch}{1.25}

\begin{table}
\centering
\begin{tabular}{ |p{2.0cm}|p{2.0cm}|p{2.0cm}|  }
\hline
TDE Name & $\sigma$ & $\log_{10} M_{\rm gal}$  \\
& $ {\rm km\, s^{-1}}$ & $\log_{10} M_\odot$  \\
\hline
 AT2017eqx & --- & $9.43^{+0.08}_{-0.08}$  \\ \hline 
 AT2018bsi & $118^{+8}_{-8}$ & $10.64^{+0.06}_{-0.06}$  \\ \hline 
 AT2019eve & --- & $9.29^{+0.17}_{-0.17}$  \\ \hline 
 AT2021ack & --- & $10.13^{+0.02}_{-0.02}$  \\ \hline 
 AT2021gje & --- & $11.00^{+0.10}_{-0.10}$  \\ \hline 
 AT2021lo & --- & $10.08^{+0.14}_{-0.14}$  \\ \hline 
 AT2022dbl & $70^{+6}_{-6}$ & $10.04^{+0.06}_{-0.06}$  \\ \hline 
 AT2022bdw & $86^{+6}_{-5}$ & $10.04^{+0.03}_{-0.03}$  \\ \hline 
 AT2022rz & --- & $9.61^{+0.17}_{-0.17}$  \\ \hline 
 AT2022dsb & --- & $10.08^{+0.02}_{-0.02}$  \\ \hline 
 AT2022hvp & --- & $10.71^{+0.04}_{-0.04}$  \\ \hline 
 AT2023clx & $61^{+2}_{-2}$ & $9.94^{+0.11}_{-0.11}$  \\ \hline 
 AT2020ksf & --- & $9.91^{+0.09}_{-0.09}$  \\ \hline 
 AT2020vdq & $44^{+3}_{-3}$ & $9.04^{+0.16}_{-0.16}$  \\ \hline 

\end{tabular}
\caption{The galactic properties of the 14 new TDEs in our sample. 
The quoted error ranges correspond to $1\sigma$ uncertainties. }
\label{gal_table}
\end{table}

\begin{table}
\centering
\begin{tabular}{ |p{2.0cm}|p{4.0cm}|  }
\hline
TDE Name & Reference  \\
\hline
 AT2017eqx & \cite{Nicholl19} \\ \hline 
 AT2018bsi & \cite{vanVelzen21} \\ \hline 
 AT2019eve & \cite{vanVelzen21}  \\ \hline 
 AT2020ksf & \cite{Wevers24}  \\ \hline 
 AT2020vdq & \cite{Somalwar23} \\ \hline 
 AT2021ack & \cite{Hammerstein21TNS} \\ \hline 
 AT2021gje &  \cite{Hammerstein21TNSb} \\ \hline 
 AT2021lo &  \cite{Yao22TNS} \\ \hline 
 AT2022dbl & \cite{Arcavi22TNS} \\ \hline 
 AT2022bdw & \cite{Arcavi22TNSb}  \\ \hline 
 AT2022rz &  \cite{Hammerstein22TNS} \\ \hline 
 AT2022hvp & \cite{Fulton22TNS} \\ \hline 
 AT2022dsb & \cite{Fulton22TNSb}  \\ \hline 
 AT2023clx & \cite{Charalampopoulos24} \\ \hline 

\end{tabular}
\caption{The discovery papers for the additional TDEs used in this work. }
\label{ref_table}
\end{table}

\label{lastpage}
\end{document}